\documentclass[12pt]{article}
\usepackage{amssymb, amsmath, amsthm}
\usepackage{lscape} 
\usepackage{mathdots}
\usepackage{thmtools}
\usepackage{mathtools}
\declaretheorem[style=definition]{example}
\usepackage{appendix}
\usepackage{bm}
\usepackage{bbm}
\usepackage{caption}
\usepackage{enumerate} 
\usepackage{graphicx}
\usepackage{natbib,hyperref}
\usepackage{setspace}
\usepackage{hyperref} 
\usepackage{color}
\usepackage{mathrsfs}
\usepackage{subfigure}
\usepackage[top=1.25in, bottom=1.25in, left=1.25in, right=1.25in]{geometry}
\newtheorem{assm}{Assumption}
\newtheorem{lem}{Lemma}
\newtheorem{prop}{Proposition}
\newtheorem{thm}{Theorem}

\newtheorem{obs}{Observation}

\begin{document}
\renewcommand{\thefootnote}{\fnsymbol{footnote}}
\renewcommand\thmcontinues[1]{Continued}
\title{Electoral Accountability and Selection with Personalized Information Aggregation}
\author{
Anqi Li\thanks{Corresponding author.} \footnote{Department of Economics, Virginia Polytechnic Institute and State University,  3126 Pamplin Hall, Blacksburg, VA 24060, United States,  angellianqi@gmail.com.}
\and Lin Hu\footnote{Research School of Finance, Actuarial Studies and Statistics, Australian National University, CBE Building 26C Kingsley Street, Canberra, ACT, 0200, Australia, lin.hu@anu.edu.au.}
}
\date{Forthcoming at Games and Economic Behavior}
\maketitle

\begin{abstract}
We study a model of electoral accountability and selection whereby heterogeneous voters aggregate incumbent politician's performance data into personalized signals through paying limited attention. Extreme voters' signals exhibit an own-party bias, which hampers their ability to discern  the good and bad performances of the incumbent. While this effect alone would undermine electoral accountability and selection, there is a countervailing effect stemming from partisan disagreement, which makes the centrist voter more likely to be pivotal. In case the latter's unbiased signal is very informative about the incumbent's performance,  the combined effect on electoral accountability and selection can actually be a  positive one.  For this reason, factors that carry a negative connotation in every political discourse---such as increasing mass polarization and shrinking attention span---have ambiguous accountability and selection effects in general.  Correlating voters' signals, if done appropriately, unambiguously improves electoral accountability and selection and, hence, voter welfare.

\bigskip


\bigskip


\bigskip

\bigskip

\end{abstract} 
\renewcommand{\thefootnote}{\arabic{footnote}}
\pagebreak

\section{Introduction}\label{sec_introduction}
Recently, the idea that tech-enabled personalized information aggregation could entail significant political consequences has been put forward in the academia and popular press \citep{sunstein, filterbubble, gentzkow}. This paper studies how personalized information aggregation by rationally inattentive voters affects electoral accountability and selection, i.e.,  society's ability to  motivate and retain talented politicians through elections.

Our premise is that rational demand for information aggregation in the digital era
is driven by limited attention capacities. As more people get  news online where the amount of available content (2.5 quintillion bytes) is vastly greater than what any individual can process in a lifetime, consumers must find ways to aggregate original content into information that is easy to process but still useful for decision-making.  Recently, this  goal has been made possible by the advent of news  aggregators, which provide customized content aggregation services  based on individuals' preferences, needs, or even personal data such as demographic and psychographic attributes, digital footprints, and social network positions.\footnote{In computing,  a news aggregator (or simply an aggregator) is a client software or a web application that aggregates syndicated web content such as online newspapers, blogs, podcasts, and vlogs in one location for easy viewing. Prominent examples of aggregators include aggregator sites, social media feeds, and mobile news apps. They operate by sifting through a myriad of online sources and displaying snippets (headline+excerpt) on their platforms. Snippets contain coarse information and do not always generate click-throughs of the original content  \citep{chioutucker}. 

Aggregators have recently gained prominence as more people get news online, from social media, and through mobile devices \citep{matsa}. The top three popular news websites in 2019: Yahoo! News, Google News, and Huffington Post, are all aggregators. The role of social media feeds in the 2016 U.S. presidential election has triggered heated debates \citep{allcottgentzkow}. See \cite{atheylocal} for a background introduction and literature survey.\label{footnote_aggregator}} Accompanying this trend is the concern that excessive reliance on aggregators  could hamper society's ability to hold elected officials accountable. As President Obama put in his farewell speech: ``For too many of us, it's become safer to retreat into our own bubbles, especially our social media feeds ... and never challenge our assumptions ... How can elected officials rage about deficits when we propose to spend money on preschool for kids, but not when we're cutting taxes for corporations? How do we excuse ethical lapses in our own party, but pounce when the other  party does the same thing? ... this selective sorting of facts; it is self-defeating.''

Our analysis is carried out in a standard model of policymaking and election.  At the outset, a candidate named $R$ assumes office and privately observes his ability, which is either high or low. A high-ability incumbent can exert high effort at a cost or low effort at no cost, whereas a low-ability incumbent can only exert low effort. Effort generates performance data, based on which voters decide whether to retain the incumbent or to replace him with a challenger named $L$ in an election. Voting is expressive as elaborated in the next paragraph, and the majority winner wins the election and earns an office rent.

To model personalized information aggregation, we depart from the representative voter paradigm and work instead with multiple voters with heterogeneous partisan preferences. Specifically, we assume that a voter's differential utility from voting for the incumbent rather than the challenger has a standard component that depends on the candidates' differential ability, as well as a new component that is captured by the voter's partisan preference parameter.  Before the election takes place, voters acquire personalized information about the incumbent's performance through paying a posterior-separable attention cost \citep{caplindean13}.  That is, each voter can aggregate the incumbent's performance data into information using a signal structure that best serves his need,  provided that the needed amount of attention for processing  information doesn't exceed his bandwidth. A voter's optimal personalized signal maximizes his expressive voting utility, subject to the aforementioned bandwidth constraint. 

Optimal personalized signals help voters form binary opinions as to which candidate to vote for. One can interpret signal realizations as voting recommendations, which voters must strictly obey. This is because any information beyond voting recommendations would only raise the attention cost without any corresponding benefit and so is wasteful. Moreover, if instead of strict obedience, a voter has a (weakly) preferred candidate that is independent of his recommendations, then he could simply vote for that candidate without paying attention, let alone exhaust his bandwidth.

We consider a symmetric environment featuring a left-wing voter, a centrist voter, and a right-wing voter. While the optimal personalized signal for the centrist voter is unbiased, that of extreme voters exhibit an own-party bias, i.e., recommend the voter's own-party candidate more often than the opposite-party candidate. Intuitively, since an extreme voter could always vote along the party line without paying attention, paying attention is useful only if he is sometimes convinced to vote across the party line. The corresponding voting  recommendation must be very strong and, in order to stay within the voter's bandwidth limit, must also be very rare (hereinafter, an \emph{occasional big surprise}), implying that the recommendation is to vote along the party line most of the time (thus an \emph{own-party bias}).  Evidence for own-party bias and occasional big surprise after the use of news aggregators has been documented in the empirical literature.\footnote{The term own-party bias refers to the positive correlation between a person's party affiliation and his propensity to support his own-party candidate. The past decade has witnessed a sharp rise in voters' own-party biases without significant changes in their intrinsic preferences \citep{fiorina, gentzkow}---a trend that could arise and persist due to the advent of personalized information aggregation technologies. 

 Occasional big surprise is a hallmark of Bayesian rationality, and its evidence is surveyed by \cite{dellavignagentzkow}.   Recently, \cite{flaxman} find that the use of news aggregators increases the own-party biases of online news consumers, as well as their opinion-intensities when supporting opposite-party candidates. \label{footnote_background}}

To illustrate how personalized information aggregation affects electoral accountability and selection, suppose  voters' population distribution is sufficiently dispersed that each voter is pivotal with a positive probability. Consider two events. In the first event, extreme voters agree on which candidate to vote for, so the \emph{incentive power} generated by their personalized signals (i.e., the ability to discern the good and bad performances of the incumbent) determines society's ability to uphold electoral accountability and selection. In the second event, extreme voters disagree about which candidate to vote for, so the centrist voter is pivotal, and the incentive power generated by his signal determines  society's ability to uphold electoral accountability and selection.  In recent years,  disagreements between extreme voters (hereinafter, \emph{partisan disagreement}) have risen sharply across a wide range of issues such as abortion, global warming, gun policy, immigration, and gay marriage \citep{fiorina, gentzkow, carroll}.

Our comparative statics exercise exploits the trade-off between the incentive power and partisan disagreement generated by extreme voters' signals. On the one hand, we find that increasing extreme voters' partisan preference parameters magnifies their own-party biases and reduces the incentive power generated by their signals. While this effect alone would undermine electoral accountability and selection, there is a countervailing effect stemming from partisan disagreement, which occurs more frequently  as extreme voters become more partisan.  The combined effect on electoral accountability and selection could be a positive one if the centrist voter's signal is very informative about the incumbent's performance. Likewise, while lowering extreme voters' bandwidths undermines the incentive power generated by their signals, it could also magnify partisan disagreement and so could potentially enhance electoral accountability and selection. Together, these results paint a nuanced picture where factors that carry a negative connotation in everyday political discourse---such as increasing mass polarization and shrinking attention span \citep{fiorina, teixeira, dunaway}---could prove conducive to electoral accountability and selection, whereas nudges designed by platforms  such as \href{https://www.allsides.com/about}{Allsides.com} to battle the rising polarization through presenting readers with balanced viewpoints could undermine electoral accountability and selection. Interestingly, correlating voters' signals, if done appropriately, unambiguously improves electoral accountability and selection, suggesting that well-conceived coordination, if not consolidation between major news aggregators, could enhance voter welfare.  

\subsection{Related literature}\label{sec_literature}
\paragraph{Rational inattention} We follow the Rational Inattention (RI) paradigm pioneered by \cite{sims1} to model information acquisition. Our voters can aggregate source data into any signal through paying a posterior separable cost.  Posterior separability has recently received attention from economists because of its axiomatic and revealed-preference foundations \citep{caplindean, tsakas, denti2022posterior,  zhong}, connections to coding theory and sequential sampling  \citep{shannon, morrisstrack, hebertwoodford}, and validations by lab experiments \citep{ambuehl2017offer, dean}. 

Our analysis exploits the flexibility of RI information acquisition, i.e., the ability to conduct any statistical experiment about a payoff-relevant state. Attentional flexibility is inherent in human decision-making \citep{dean, sali2020neural}, and its impacts on bargaining, incentive contracting, and other branches of economics are surveyed by \cite{risurvey}.  Recent papers by  \cite{own}, \cite{matejka}, and \cite{yuksel}, study electoral competition with flexible information acquisition. In \cite{own}, information is aggregated by an attention-maximizing infomediary for arbitrary coalitions of voters. Here, individual voters can aggregate information optimally themselves as in the standard RI paradigm.

\paragraph{Rational ignorance} The idea that costly information acquisition could shape political outcomes has a long history that dates back to \cite{downs}. Yet until recently, the predominant way that political economists model information acquisition (commonly known as \emph{rational ignorance}) imposes severe limitations on the  experiments that decision-makers can conduct. Among others,  \cite{persico} restricts the set of feasible experiments to a singleton, and \cite{martinelli} allows voters to acquire only unbiased signals. 

The only exceptions are studies on filtering bias, which advocate the idea that even rational consumers can exhibit a preference for biased information when constrained by information processing capacities (see, e.g., \citealp{calvert, burke, suen, oliveros, che}). While these studies predict an own-party bias and, implicitly, an occasional big surprise as we do, they work with ad-hoc information acquisition technologies that lack axiomatic and coding theoretic foundations. By showing that qualitatively similar predictions can be made from studying RI information acquisition, we provide foundations for these predictions and demonstrate the potential of RI as a workhorse model for studying costly and yet flexible information acquisition in political economy problems. When it comes to aggregate outcomes and detailed comparative statics, our model generates new predictions that set itself apart from traditional models; see Footnote \ref{footnote_nonmonotonicity} for  details.\footnote{As another example, consider the information aggregation technology studied by  \cite{suen}, which partitions realizations of a continuous state variable into two cells using a threshold rule. Since the resulting signal realizations are monotonic in voters' partisan preferences (i.e., if a left-wing voter is recommended to vote for candidate $R$, then a right-wing voter must receive the same recommendation),  median voter's signal determines the election outcome despite a pluralism of voters and media outlets. }

\paragraph{Electoral accountability and selection} Most existing studies on electoral accountability and selection work with a representative voter who faces either an exogenous information environment or self-interested media with persuasion motives (see, e.g., \citealp{besleyprat, ashworthdemesquita14, amanda,  wolton}). \cite{egorov} and \cite{pratsurvey}  study accountability models with heterogeneous voters as we do, although their analysis abstracts away from  information acquisition and examines different voting games from ours.\footnote{Specifically, \cite{egorov} examines a retrospective voting model with pure moral hazard but no adverse selection. In \cite{pratsurvey}, equilibrium policy  depends only on the population of informed voters  but not on the disagreement between different voter groups. } There are also accountability models with rational ignorance, all featuring a single voter and non-RI, rigid, information acquisition (see, e.g., \citealp{svolik, trombetta}). 

\paragraph{Common agency} The theory of common agency games with moral hazard pioneered by \cite{bernheimwhinston} has been widely applied to the studies of bureaucracy, international trade, etc. We study a special, albeit new case of the general framework proposed by \cite{peters}, whereby the allocation offered by principals (here, voters) to the agent (here, the incumbent) consists of signal structures that monitor the latter's performance. \cite{fahad} also study a common agency game with endogenous monitoring, although their financial contracting problem differs completely from ours, their principals are homogeneous, and their monitoring technologies are stylized. 
 
\section{Model}\label{sec_model}

\subsection{Setup}
There is an incumbent named $R$, a challenger named $L$, and three voters $k \in \mathcal{K}=\left\{-1,0,1\right\}$ of a unit total mass. Voter $k$'s mass is  $f_k \in \left(0,1\right)$, and his partisan preference between the two candidates is captured by $v_k \in \left(-1,1\right)$. The game begins with the incumbent assuming office and privately observing his ability $\theta$, which is either high ($\theta=h$) or low ($\theta=l$) and has zero mean, i.e., $h>0>l$ and $\mathbb{E}[\theta]=0$. A high-ability incumbent can exert high effort ($a=1$) at a cost $c>0$ or low effort ($a=0$) at no cost, whereas a low-ability incumbent can only exert low effort. The incumbent's effort choice $a \in \left\{0,1\right\}$ is his private information, and it generates performance data modeled as a finite random variable with support $\Omega$ and p.m.f. $p_{a}$ (more on this later). After that, an election takes place, in which voters decide whether to re-elect the incumbent  or to replace him with the challenger whose expected ability is normalized to zero. The election outcome is determined by simple majority rule with ties broken in favor of the incumbent, and the winning candidate earns one unit of office rent. 

Voting is \emph{expressive}, meaning that voters care about their individual voting decisions but not about the aggregate voting outcome.\footnote{See \cite{matejka} for a recent discussion about the motives behind expressive voting, e.g., citizen duty, the desire to inform others about one's opinion and decision, etc. } Voter $k$'s differential utility $v_k+\theta$ from choosing the incumbent over the challenger equals his partisan preference parameter $v_k$,  plus the incumbent's ability $\theta$ relative to the challenger's. The first part of the utility is new, whereas the second part is standard and captures the quality of future policy-making.  

Before the election takes place, voters can acquire personalized information about the incumbent's performance. A \emph{signal structure} (or simply a   \emph{signal}) is a mapping $\Pi: \Omega \rightarrow \Delta\left(\mathcal{Z}\right)$, where each $\Pi\left(\cdot \mid \omega\right)$, $\omega \in \Omega$, specifies a probability distribution over a finite set $\mathcal{Z}$ of signal realizations conditional on the incumbent's performance state being $\omega$. Processing the information generated by $\Pi$ incurs an \emph{attention cost} $I\left(\Pi\right)$ (more on this later), which must not exceed voter $k$'s \emph{bandwidth} $I_k>0$ in order for him to find the signal structure feasible. After that, the voter observes the signal realization, updates his belief about the incumbent's performance and  ability, and casts his vote. 

The game sequence is summarized as follows.
\begin{enumerate}
\item The following events occur simultaneously:
  \begin{enumerate}[(a)]
\item the incumbent observes his ability and makes an effort choice; effort generates performance data; 
\item voters specify personalized signal structures. 
\end{enumerate}
\item Voters observe signal realizations and cast votes.
\end{enumerate}

We consider a \emph{symmetric} environment where $v_{-k}=-v_k$, $f_{-k}=f_{k}$, and $I_{-k}=I_{k}$ $\forall k \in \mathcal{K}$. Voter $-1, 0, 1$'s partisan preference parameters satisfy $v_{-1}<0, v_0=0$, and $v_1>0$, and they are called \emph{left-wing}, \emph{centrist}, and \emph{right-wing}, respectively. For the most part, we assume that signals are \emph{conditionally independent} across voters; see, however, Section \ref{sec_extension} for an extension to correlated signals.  

Our solution concept is \emph{perfect Bayesian equilibrium}, or \emph{equilibrium} for short. We say that an equilibrium \emph{sustains accountability} if it induces the high-ability incumbent to exert high effort, and that  \emph{accountability is sustainable} if it can be sustained in an equilibrium. We also define the degree of \emph{electoral  selection} as the expected ability of the elected official at the end of the game. Since voters care only about the ability of the elected official, their equilibrium expected utilities equal their partisan preference parameters plus the degree of electoral selection. Accountability benefits voters by helping the high-ability incumbent signal his type.

 Our main research question concerns the accountability and selection effects of personalized information aggregation. Throughout the paper, we assume that whenever the incumbent is indifferent between exerting high and low effort, he exerts high effort. 

\subsection{Key assumptions}\label{sec_discussion}

\paragraph{Performance data} Our assumptions about the performance data are twofold. 

\begin{assm}\label{assm_singledimensional}
(i) $\omega=\mathbb{E}[\theta \mid \omega, a_h=1]$ $\forall \omega \in \Omega$; (ii) $\Omega=\{-1,1\}$. 
\end{assm}

Assumption \ref{assm_singledimensional} has two parts. Part (i) of the assumption is without loss of generality (w.l.o.g.).\footnote{To see why, note that our game always has an equilibrium without accountability,  whereby the high-ability incumbent exerts low effort, and voters best respond by acquiring no information about the incumbent's performance. 

 Meanwhile in any equilibrium with accountability, voters must best respond to the high-ability incumbent's equilibrium effort choice, which we denote by  $a_h=1$. Let us begin with an arbitrary finite probability space $(\Omega, p_a)$. Since voters care only about the incumbent's expected ability $\omega' \coloneqq \mathbb{E}[\theta \mid \omega, a_h=1]$, their best-responding signal structures must depend \emph{only} on $\omega'$, i.e., $\Pi: \{\omega'\} \rightarrow \Delta(\mathcal{Z})$, by \cite{mckay}. Anticipating this,  the high-ability incumbent makes an optimal effort choice based on $(\Omega', p_0', p_1')$, where  $\Omega'\coloneqq \{\omega'\}$ and $p_a'(\omega')\coloneqq p_a\{\omega \in \Omega: \mathbb{E}[\theta\mid \omega, a_h=1]=\omega'\}$. Taken together,  we conclude that the tuple $(\Omega', p_0', p_1')$ contains all the information we as modelers need to determine whether accountability is sustainable or not.  Finally, relabel every $\omega$ such that $\mathbb{E}[\theta\mid \omega, a_h=1]=\omega'$
 as ``$\omega'$''. Then $\mathbb{E}[\theta \mid \omega', a_h=1]=\omega'$ $\forall \omega' \in \Omega'$ by construction. 
 
 For the above reasons, we shall hereinafter work only with the probability space $(\Omega', p_a')$ induced by $(\Omega, p_a)$, and assume w.l.o.g. that $\omega'=\mathbb{E}[\theta\mid \omega', a_h=1]$ $\forall \omega' \in \Omega'$.  \label{footnote_assumption} } Part (ii) of it  serves two instrumental roles: (a) restrict our attention to the case of binary performance data (for ease of analysis and interpretation); (b) help ensure that the environment is symmetric in any equilibrium with accountability. Formally, 

\begin{obs}\label{obs_omega}
The following statements are true under Assumption \ref{assm_singledimensional}(ii). 
\begin{enumerate}[(i)]
\item We can interpret $\omega=1$ as \emph{good performance} and $\omega=-1$ as \emph{bad performance}, since the likelihood of generating $\omega=1$ rather than $\omega=-1$ increases with the incumbent's effort, i.e., 
$\frac{p_1\left(1\right)}{p_1\left(-1\right)}>\frac{p_0\left(1\right)}{p_0\left(-1\right)}$.
\item Each $\omega \in \Omega$ is realized with $.5$ probability in case  $a_h=1$.
\end{enumerate}
\end{obs}

The proof of Observation \ref{assm_singledimensional} is relegated to Appendix \ref{sec_proof_observation}. An important step of the proof is to delineate the restrictions that Assumption \ref{assm_singledimensional}(ii), together with the normalization $\mathbb{E}[\theta]=0$, imposes on the model primitives. 

\paragraph{Personalized information aggregation} Our analysis is inspired by two stylized facts: (i) the need for information aggregation in today's digital age is driven by limited attention capacities; (ii) recent technology advances make personalized information aggregation possible.  We capture these facts by following the RI paradigm, i.e.,  allow voters to acquire any finite signal about the incumbent's performance through paying a posterior-separable attention cost. 

A  signal structure (or simply a signal) governs how performance data are aggregated into signal realizations. Let $\Pi: \Omega\rightarrow \Delta\left(\mathcal{Z}\right)$ be any finite signal structure, where each $\Pi(\cdot \mid \omega)$, $\omega \in \Omega$,  specifies the probability distribution over a finite set $\mathcal{Z}$ of signal realizations  conditional on the true performance state being $\omega$. In the case where $a_h=1$, each $z\in \mathcal{Z}$ occurs with probability 
\[
\pi_z=\sum_{\omega \in \Omega} \Pi\left(z \mid \omega\right)/2, 
\]
and \[
\mu_z=\sum_{\omega \in \Omega}\omega \cdot \Pi\left(z \mid \omega\right)/\left(2\pi_z\right)
\]
is the posterior mean of the incumbent's performance and, hence,  his ability, conditional on the signal realization being $z$. The tuple $(\pi_z, \mu_z)_{z \in \mathcal Z}$ fully captures the distribution over posterior beliefs that is induced by the consumption of $\Pi$. 

Next are our key assumptions about the cost of paying attention. 

\begin{assm}\label{assm_attention}
When $a_h=1$, the attention cost associated with acquiring $\Pi: \Omega \rightarrow \Delta(\mathcal{Z})$ is 
\begin{equation}\label{eqn_attentionfunction}
I\left(\Pi\right)= \sum_{z\in \mathcal{Z}} \pi_z \cdot h\left(\mu_z\right). 
\end{equation}
The function $h: \left[-1,1
\right]\rightarrow \mathbb{R}_+$ is continuous on $\left[-1,1\right]$ and twice differentiable on $\left(-1,1\right)$. It satisfies (i) strict convexity on $\left[-1,1\right]$ and $h\left(0\right)=0$; (ii) symmetry around zero; (iii) $h\left(1\right)>I_k$ $ \forall k \in \mathcal{K}$.
 \end{assm}

Equation (\ref{eqn_attentionfunction}) coupled with Assumption \ref{assm_attention}(i) is equivalent to \emph{weak posterior separability} (WPS)---a notion proposed by \cite{caplindean13} to generalize Shannon entropy as a measure of attention cost. For a review of the foundations for WPS, see Section \ref{sec_literature}.\footnote{The connection between posterior separability and sequential sampling is noteworthy. According to \cite{shannon},  \cite{morrisstrack}, and  \cite{hebertwoodford}, one can think of a snippet as a small piece of information encountered by a decision-maker before he stops acquiring information. Under general conditions, the expected number of snippets consumed by the decision-maker is the posterior-separable attention cost that is needed for implementing a signal structure.} In the current setting, WPS  stipulates that consuming a null signal requires no attention, and that more attention is needed for moving one's posterior belief closer to the true state, and as the signal becomes more Blackwell-informative. The high-level idea is that attention is a scarce resource that helps reduce uncertainties about the incumbent's performance and, ultimately, his ability. 

Parts (ii) of Assumption \ref{assm_attention} imposes symmetry on our problem. It is satisfied by commonly used attention functions in the RI literature, including  reductions in the variance and Shannon entropy of the state before and after information consumption.\footnote{
 $h\left(\mu\right)=\mu^2$ and Binary entropy function $\left(\left(1+\mu\right)/2\right)$, respectively, in these cases.}
 
  Part (iii) of Assumption \ref{assm_attention} creates a role for information aggregation, saying that voters must garble performance data in order to stay within their bandwidth limits.\footnote{An alternative, standard assumption made  in the RI literature is that decision-makers pay a constant unit cost for acquiring information.  The  comparative statics results generated by this alternative model  can be decomposed into those generated by the current model.  For example,  varying a voter's partisan preference parameter (i) changes the total amount of attention he pays,  and (ii) adjusts the location of his signal structure on the attention level curve.  The current analysis examines these basic effects in isolation.   Since these effects tend to move in different directions, the combined effect is in general ambiguous. } Without this assumption,  information aggregation becomes trivial, a topic we will turn to in Section \ref{sec_benchmark}.

\section{Analysis}\label{sec_analysis}
This section examines the accountability and selection effects of personalized information aggregation.  We begin with a benchmark case where voters face no bandwidth limit in Section \ref{sec_benchmark}.  In Sections \ref{sec_aggregator} and \ref{sec_accountability}, we reintroduce bandwidth limits  and give equilibrium characterizations. Section \ref{sec_cs} investigates equilibrium comparative statics. 

Three remarks before we proceed. First,  all upcoming  results exploit Assumptions \ref{assm_singledimensional} and  \ref{assm_attention} unless otherwise specified. Second, it is easy to show that whenever an accountability equilibrium exists, it is unique. Third, due to symmetry, we can and will use $v_1$ and $I_1$ to represent extreme voters' partisan preference parameter and bandwidth, respectively. 

\subsection{Benchmark}\label{sec_benchmark}
In this section, we lift voters' bandwidth limits so that they can process the incumbent's performance data without error. The next lemma examines when accountability can arise in an equilibrium in this benchmark case.

\begin{lem}\label{lem_benchmark}
Let everything be as in Assumptions \ref{assm_singledimensional} and \ref{assm_attention} except that $I_k\geq h\left(1\right)$ $\forall k \in \mathcal{K}$. Then (i) all voters vote for the incumbent if $\omega=1$, and for the challenger if $\omega=-1$, in case $a_h=1$. (ii) Accountability is sustainable if and only if $1 \geq \hat{c}$,
where $1$ is the differential probability that the incumbent wins re-election in states $\omega=1$ and $\omega=-1$, given voters' behaviors in Part (i), and 
\[
\hat{c} = \frac{c}{p_1\left(1\right)-p_0\left(1\right)}
\]
represents the threshold that must be crossed in order to sustain accountability in an equilibrium. (iii) The degree of electoral selection equals $1/2$ with accountability and zero without accountability.
\end{lem}

\begin{proof}
Part (i) of the lemma holds because voters have mild partisan preferences, i.e., $v_k-1<0<v_k+1 $ $\forall k$. Under this assumption, voting for the incumbent if $\omega=1$ and for the challenger if $\omega=-1$ is a best response to $a_h=1$ for all voters.\footnote{If $|v_{\pm 1}|>1$ instead, then extreme voters would always vote along party lines regardless of the incumbent's performance, which brings us back to the representative voter paradigm where centrist voter's vote determines the election outcome. For this reason, adding voters with $|v|>1$ to the model wouldn't affect the upcoming analysis. }

 To demonstrate Part (ii) of the lemma, note that if voters behave as in Part (i), then a high-ability incumbent's winning probability is $p_1(1)$ if he exerts high effort, and $p_0(1)$ if he exerts low effort. Exerting high effort is optimal if and only if $p_1(1)-p_0(1) \geq c$, or equivalently $1 \geq \hat{c}$. 
 
 Turning to Part (iii), note that the degree of electoral selection is $p_1(1) \cdot 1+ p_1(-1)\cdot 0=1/2$ in an accountability equilibrium. Without accountability, performance data carry no information about the incumbent's ability, so the degree of electoral selection is $\mathbb{E}[\theta]=0$.
\end{proof}

\subsection{Optimal personalized signal}\label{sec_aggregator}
In this section, we bring back  bandwidth limits and solve for the signal structures that maximize voters' expressive voting utilities in case $a_h=1$ (hereinafter, \emph{optimal personalized signals}).  

Our starting observation is that extreme voters could always vote along party lines without paying attention. For these voters, paying attention is beneficial only if they are sometimes convinced to vote across party lines.  After absorbing the information generated by $\Pi$, voter $k$ strictly prefers candidate $R$ to candidate $L$ if $v_k+\mu_z>0$, and he strictly prefers candidate $L$ to candidate $R$ if $v_k+\mu_z<0$. Ex-ante, voter $k$'s expected utility gain from consuming $\Pi$ is 
\[
V_k\left(\Pi\right)=
\sum_{z \in \mathcal{Z}} \pi_z \cdot \nu_k\left(\mu_z\right)
\]
where 
\[\nu_k\left(\mu_z\right)=
\begin{cases}
\left[v_k+\mu_z\right]^{+} & \text{ if } k \leq 0,\\
-\left[v_k+\mu_z\right]^{-} & \text{ if } k>0.
\end{cases}\]
An optimal personalized signal for voter $k$ thus solves 
\[
\max_{\mathcal{Z}, \Pi: \Omega \rightarrow \Delta\left(\mathcal{Z}\right)} V_k\left(\Pi\right) \text{ s.t. } I\left(\Pi\right) \leq I_k.
\]

The next lemma provides preliminary characterizations for optimal personalized signals.  

\begin{lem}\label{lem_binary}
In the case where $a_h=1$, the optimal personalized signal for any voter $k \in \mathcal{K}$, hereinafter denoted by $\Pi_k$, is unique, exhausts his bandwidth, and prescribes voting recommendations that he strictly obeys, i.e., $I\left(\Pi_k\right)=I_k$, $\mathcal{Z}_k=\left\{L, R\right\}$, and $v_k+\mu_{L, k}<0<v_k+\mu_{R,k}$. 
\end{lem}

The idea behind Lemma \ref{lem_binary} is straightforward: Since a voter's ultimate goal is to choose between the candidates, any information beyond voting recommendations would only raise his attention cost without any corresponding benefit and so is wasteful. Voting recommendations must be strictly obeyed, because if a voter has a (weakly) preferred candidate that is independent of his recommendations, then he could always vote for that candidate without paying attention,  let alone exhaust his bandwidth.

In light of Lemma \ref{lem_binary}, we shall hereinafter focus on signal structures that prescribe voting recommendations to voters. Call a signal structure \emph{neutral} if it recommends both candidates with equal probability, \emph{$L$-biased} if it recommends candidate $L$ more often than candidate $R$, and \emph{$R$-biased} if it recommends candidates $R$ more often than candidate $L$. The next lemma characterizes the biases of optimal personalized signals.

\begin{lem}\label{lem_skewness}
In the case where $a_h=1$, the optimal personalized signal for the centrist voter is neutral, whereas that of an extreme voter exhibits an  \emph{own-party bias}, i.e., the signal recommends the voter's own-party candidate more often than his opposite-party candidate, and it does so more often as the voter's partisan preference parameter increases. 
\end{lem}

To develop intuition, note that  an extreme voter prefers his own-party candidate ex ante. To satisfy strict obedience,  the recommendation to vote across the party line must be very strong and, in order to stay within the voter's  bandwidth limit, must also be very rare (hereinafter an \emph{occasional big surprise}). Most of the time, the recommendation is to vote along the party line (thus an own-party bias).  Evidence for own-party bias and occasional big surprise has already been discussed in Footnote \ref{footnote_background}. 

To facilitate analysis of the accountability effect of personalized information aggregation, we introduce two concepts. The first concept is called the \emph{incentive power} generated by an optimal personalized signal $\Pi_k$. It is defined as the differential probability that $\Pi_k$ recommends the incumbent for re-election in good and bad performance states:
\[P_k=\Pi_k\left(R \mid \omega=1\right)-\Pi_k\left(R \mid \omega=-1\right).\]
Intuitively, $P_k$ captures voter $k$'s ability in discerning   good and bad performance states. In the representative voter paradigm, it determines the voter's ability to hold the incumbent accountable.

The second concept: \emph{partisan disagreement}, is defined as the  probability that extreme voters receive conflicting voting recommendations and so disagree about which candidate to vote for:
\[D=\mathbb{P}_{\Pi_{\pm 1}}\left(z_{1} \neq z_{-1} \mid a_h=1\right).\]
The past decade has witnessed sharp rises in partisan disagreement across a wide range of issues, including  abortion, global warming, gun policy, immigration, and gay marriage. 
In the current setting, $D>0$ because extreme voters' signals  recommend their opposite-party candidates with positive, albeit small, probabilities. Such  occasionally big surprises play an implicit yet crucial role in the upcoming analysis. 

The next lemma establishes the tensions between incentive power and partisan disagreement as we vary model primitives. 
\begin{lem}\label{lem_variable}
\begin{enumerate}[(i)]
\item As extreme voters become more partisan, the incentive power generated by their optimal personalized signals decreases, and partisan disagreement arises more frequently, i.e., $P_{-1}$ and $P_1$ are decreasing in $v_1$, whereas $D$ is increasing in $v_1$.
\item As voters' bandwidths increase, the incentive power generated by their optimal personalized signals increases, whereas partisan disagreement may arise more frequently or less frequently, i.e., $P_k$ is increasing in $I_k$ for any $k \in \mathcal{K}$, whereas $D$ is in general non-monotonic in $I_1$.
\end{enumerate}
\end{lem}

Part (i) of Lemma \ref{lem_variable} holds because as extreme voters become more partisan, they endorse their own-party candidates more often, regardless of the true performance state. 

Part (ii) of the lemma holds because as voters' bandwidths increase, their personalized signals become more Blackwell informative and so generate more incentive power individually.\footnote{This result is reminiscent of Proposition 5.2 of \cite{dewatripont} (DJT), which shows that garbling the market signal in career-concern models undermines agent's incentive to exert costly effort under general conditions. In both DJT and the single-voter version of the  current paper, there is a one-to-one mapping between the signal realization and the  agent's outcome (market wage in DJT, election outcome here). In accountability models such as the one studied by \cite{ashworthdemesquita14}, the mapping isn't  one-to-one (e.g., signal is normally distributed whereas voting decision is binary), which explains why increased signal informativeness could have ambiguous incentive effects on the agent. \label{footnoteDJT}  }  Partisan disagreement may arise more frequently or less frequently. The first situation happens if, due to the flexibility in attention allocation, extreme voters become significantly more supportive of their own-party candidates when performance data are favorable, but they do not cut back support as much when performance data are unfavorable. Condition (\ref{eqn_Dlambda}) in Appendix \ref{sec_proof_lemma}---which exploits the curvature of the attention cost function---is sufficient and necessary for $D$ to decrease with $I_1$. Examples that satisfy and violate the condition are presented right after its statement. 

\subsection{Electoral accountability and selection}\label{sec_accountability}
In this section, we take the optimal personalized signals solved in the previous section as given, and verify whether they can induce the high-ability incumbent to exert high effort. If the answer to this question is positive, then accountability can be sustained in an equilibrium. We also solve for the equilibrium degree of electoral selection with and without accountability.

Our analysis exploits a key concept called  \emph{societal incentive power}, defined as the differential probability that the incumbent wins re-election in good and bad performance states, given the signal structures solved in the previous section:  
\[\xi=\mathbb{P}_{\Pi_{k}s}\left(R \text { wins re-election} \mid \omega=1\right)-\mathbb{P}_{\Pi_{k}s}\left(R \text { wins re-election} \mid \omega=-1\right).\]
Intuitively, $\xi$ captures society's ability to  uphold electoral accountability and selection through rewarding good performances and punishing bad performances. The next theorem solves for $\xi$ and expresses equilibrium outcomes as functions of $\xi$.

\begin{thm}\label{thm_accountability} The societal incentive power equals 
\[
\xi=\begin{cases}
P_0 &\text{ if } f_0 \geq 1/2,\\
P_1+DP_0& \text{ if } f_0 <1/2.
\end{cases}
\]
Accountability is sustainable if and only if $\xi \geq \hat{c}$. The degree of electoral selection equals $\xi/2$ with accountability and zero without accountability. 
\end{thm}

We distinguish between two cases: $f_0 \geq 1/2$ and  $f_0<1/2$. In the first case, the centrist voter is the only pivotal voter, so the incentive power generated by his personalized signal determines the societal incentive power.  In the second case, each voter is pivotal with a positive probability, so there are two events to consider. In the first event,  extreme voters reach a consensus as to which candidate to vote for, so the centrist voter is non-pivotal. The societal incentive power stemming from this event is simply the incentive power generated by extreme voters'  signals. In the second event, extreme voters disagree about which candidate to vote for, so the centrist voter is pivotal. The societal incentive power stemming from this event equals its probability times the incentive power  generated by the centrist voter's signal.

As $\xi$ increases, the society as a whole becomes better at rewarding good performances and punishing bad performances. Accountability becomes sustainable the moment when $\xi$ crosses the threshold $\hat{c}$ from below. As for selection, notice that without accountability,
performance data carry no information about the incumbent's ability, which makes selection impossible. With accountability, the degree of electoral selection is proportional to $\xi$.  

\subsection{Comparative statics}\label{sec_cs}
This section examines the comparative statics of societal incentive power.  All results are built upon, and sometimes follow immediately from Lemma \ref{lem_variable} and Theorem \ref{thm_accountability}. 

Our first proposition examines the effect of varying  extreme voters' partisan preference parameter on  societal incentive power.

\begin{prop}\label{prop_skewness}
In the case where $f_0<1/2$, the societal incentive power $\xi=P_1+DP_0$ is in general non-monotonic in extreme voters' partisan preference parameter $v_1$.
\end{prop}

Proposition \ref{prop_skewness} showcases the tension between incentive power and partisan disagreement as we vary extreme voters' partisan preference parameter. As extreme voters become more partisan, their personalized signals become more biased and generate less incentive power individually---recall Lemma \ref{lem_variable}. While this effect alone would make accountability harder to sustain, there is a countervailing effect stemming from partisan disagreement, which arises more frequently as extreme voters become more partisan. In case of disagreement, the centrist voter is pivotal, and his contribution $P_0$ to the societal incentive power increases with his bandwidth $I_0$. Consider three cases. 
\begin{itemize}
\item In one extreme situation where $I_0\approx 0$, $P_0 \approx 0$, i.e., centrist voter can barely distinguish the good and bad performances of the incumbent. In that situation, the societal incentive power depends mainly on the incentive power generated by extreme voters' signals, i.e., $\xi \approx  P_1$. The last term is decreasing in $v_1$ by Lemma \ref{lem_variable}.
\item  In another extreme situation where $I_0 \approx 1$,  centrist voter can process the incumbent's performance data with few errors and so can distinguish the good and bad performances of the incumbent almost perfectly. In that situation,  $P_0\approx 1$ and $\xi\approx P_1+D$, where the last term is maximized when $v_1$ is sufficiently large (see Appendix \ref{sec_proof_theorem} for a formal proof). 
\item For in-between cases, $\xi$ can vary non-monotonically with $v_1$ as demonstrated by  the next example. 
\end{itemize} 

\begin{example}\label{exm_monotone}
In Appendix \ref{sec_proof_observation}, we demonstrate that any binary signal structure $\Pi: \Omega \rightarrow \Delta(\{L, R\})$ can be represented by the profile $(\mu_L, \mu_R)$ of the posterior means it induces. A signal structure is neutral if $|\mu_L|=\mu_R$, $L$-biased if $|\mu_L|<\mu_R$, and $R$-biased if $|\mu_L|>\mu_R$. In  the case of quadratic attention cost, i.e., $h\left(\mu\right)=\mu^2$, a typical level curve of the attention cost function takes the form of
 \[\mathcal{C}(I)= \left\{\left(\mu_L, \mu_R\right) \in [-1,0]\times [0,1]: I(\mu_L,\mu_R)=|\mu_L|\mu_R=I\right\}.\] Define $z=(|\mu_L|+I_1/|\mu_L|)^{-1}$. Then $P_1$, $D$, and $\xi$ can be expressed as linear and quadratic functions of $z$: 
 \[P_1=2I_1z,  D=1-2I_1\left(1+I_1\right) z^2, \text{ and } \xi=-2P_0I_1\left(1+I_1\right)z^2+2I_1 z+P_0.\] 
 
 As we increase $v_1$ from zero to infinity,  right-wing voter's personalized signal traverses along his attention level curve $\mathcal{C}(I_1)$ from its neutral element $\left(\sqrt{I_1}, \sqrt{I_1}\right)$ to its most $R$-biased element $\left(1, I_1\right)$ as depicted in Figure  \ref{figure1}. 
 \begin{figure}
\centering
  \begin{subfigure}
  \centering
     \includegraphics[width=.35\textwidth]{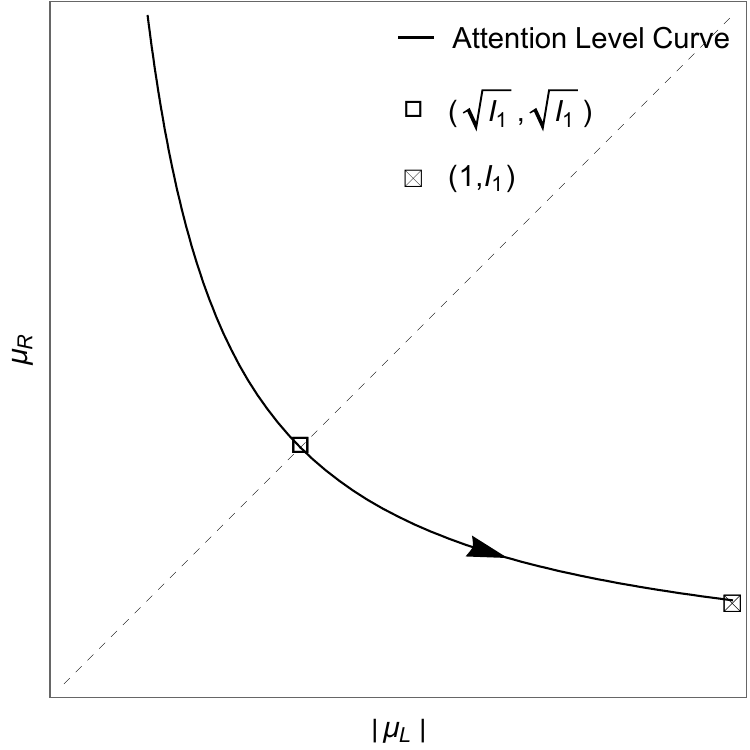}
 \label{fig_ex11}
    \end{subfigure}    \quad \quad \quad \quad 
    \begin{subfigure}
    \centering
     \includegraphics[width=.35\textwidth]{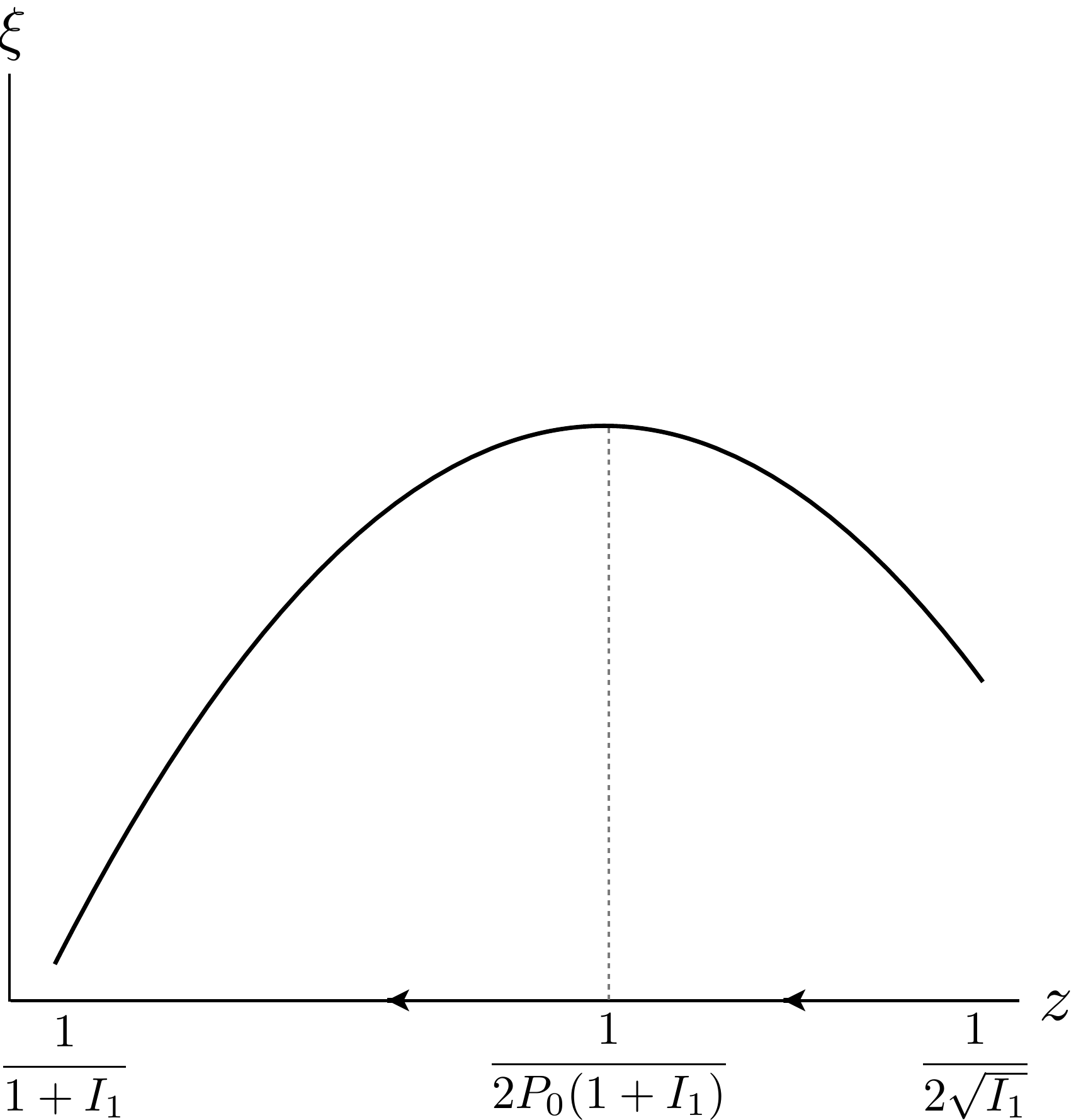}
    \label{fig_ex12}
\end{subfigure}
\caption{The left panel depicts how right-wing voter's personalized signal traverses along his attention level curve from $\left(\sqrt{I_1}, \sqrt{I_1}\right)$ to $\left(1, I_1\right)$ as $v_{1}$ increases. The right panel plots $\xi$ against $z$ for the case where $2 P_0 \left(1+I_1\right) \in (2\sqrt{I_1}, 1+I_1)$.}\label{figure1}
\end{figure} 
During that process, $z$ decreases from $(2\sqrt{I_1})^{-1}$ to $(1+I_1)^{-1}$, so $P_1$ decreases whereas $D$ increases. The overall effect on $\xi$ depends on parameter values. The case where $2 P_0 \left(1+I_1\right) \in (2\sqrt{I_1}, 1+I_1)$ is the most interesting, as neither the incentive power effect nor the partisan disagreement effect dominates the other in that case. As we decrease $z$ as above, $\xi$ first increases and then decreases as depicted in Figure \ref{figure1}. 
 $\hfill \diamondsuit$
\end{example} 

In recent years, news aggregators such as \href{https://www.allsides.com/about}{Allsides.com} have been designed and built to battle the rising polarization  through presenting readers with balanced viewpoints. The current analysis casts doubts on the usefulness of these aggregators, as feeding extreme voters with unbiased signals is mathematically equivalent to reducing $v_1$,  and so could make electoral accountability and selection harder, not easier, to sustain according to  Proposition \ref{prop_skewness}. 

Our second proposition examines how changing  voters' bandwidths affects the societal incentive power. 

\begin{prop}\label{prop_bandwidth}
In the case where $f_0<1/2$, the societal incentive power $\xi=P_1+DP_0$ is increasing in the centrist voter's bandwidth $I_0$ but is in general non-monotonic in extreme voters' bandwidth $I_1$. 
\end{prop}

The level of political knowledge among ordinary citizens is viewed by political scientists as an important determinant for how well elected officials can be held accountable (see  \citealp{ashworthdemesquita14} and the references therein). Recently, scholars and pundits have voiced growing concerns over people's shrinking attention span that results from an overabundance of entertainment, the advent of the Internet and mobile devices, and the intensified  competition between firms for consumer eyeballs \citep{teixeira, dunaway}. Proposition \ref{prop_bandwidth} paints a rosier  picture. As a voter's bandwidth decreases, his personalized signal generates less incentive power, which would undermine electoral accountability and selection in the representative voter paradigm. Yet there is an additional effect that stems from partisan disagreement, which may arise more frequently as extreme voters' bandwidth increases. Thus,  while lowering centrist voter's bandwidth unambiguously undermines electoral accountability and selection, nothing as clear-cut can be said about extreme voters' bandwidth.

Our last proposition examines how the societal incentive power varies with voters' population distribution. 

\begin{prop}\label{prop_population}
Let $\xi$ and $\xi'$ be the societal incentive power under two population distributions $f$ and $f'$ such that $f_0 \geq 1/2>f'_0$. Then $\xi'-\xi=P_1-\left(1-D\right)P_0$ is decreasing in the centrist voter's bandwidth $I_0$ but is in general non-monotonic in extreme voters' partisan preference parameter $v_1$ and bandwidth $I_1$.  
\end{prop}

Recently, a growing body of the literature has been devoted to understanding voter polarization (also termed \emph{mass polarization}).  Notably, \cite{fiorina} define mass polarization as a bimodal distribution of voters'  preferences on a liberal-conservative scale, and \cite{gentzkow} develops a related concept that measures the average ideological distance between Democrats and Republicans. Inspired by these authors, we define \emph{increasing mass polarization}  as a mean-preserving spread of  voters' partisan preferences. Proposition \ref{prop_population} shows that  increasing mass polarization could prove conducive to electoral accountability and selection despite its negative connotation in everyday discourse: As we keep redistributing voters' population from the center to the margin, extreme voters will eventually become pivotal with a positive probability.   From that moment onwards,  they  contribute to the societal incentive power, whereas  centrist voter's contribution is discounted by the probability that a consensus is reached among extreme voters. The overall effect on electoral accountability and selection is in general non-monotonic in extreme voters' partisan preference parameter and bandwidth according to Lemma \ref{lem_variable}. 

\section{Extensions}\label{sec_extension}
\paragraph{Continuous performance states} Several findings are noteworthy when we extend the analysis to a continuum of performance states. First, optimal personalized signals remain binary and still exhibit an own-party bias and occasional big surprise for the same reason as before. Due to space limitations,  we won't prove these results formally here, but refer interested readers to Proposition O5,  Online Appendix O.3 of \cite{own} for details.\footnote{Both the problems studied here and in Online Appendix O.3 of \cite{own} can be reduced to a standard rational inattention problem: $\max_{\Pi} V_k(\Pi)-\lambda_k I(\Pi)$, where $V_k(\cdot)$ is the utility gain from information consumption, and $I(\Pi)$ is the attention cost. The main difference lies in $\lambda_k$, which represents the Lagrange multiplier associated with the bandwidth constraint $I_k \geq I(\Pi)$ here, but that of a different constraint in HLS. Despite this difference, the solutions to the two problems are qualitatively the same.   }

Second,  with a continuum of states, the societal incentive power can vary non-monotonically with model primitives for two additional reasons. The first reason is mechanical, namely the decomposition of societal incentive power into features of  individual voters' signals (as in Theorem \ref{thm_accountability}) becomes less straightforward. The second reason is more interesting: due to the flexibility in allocating one's attention across a continuum of states, even the incentive power generated by an individual voter's signal can vary non-monotonically with his partisan preference parameter; see Appendix \ref{sec_state} for details. 

\paragraph{Correlated signals} So far we have restricted signals to be conditionally independent across voters. By choosing the right correlation structure, holding  marginal signal distributions fixed, we can always weakly improve, and sometimes strictly improve the societal incentive power; see Appendix \ref{sec_correlation} for a numerical example. This finding suggests that well-conceived coordination, if not consolidation between major news aggregators, could enhance electoral accountability and selection and, hence, voter welfare.

\section{Concluding remarks}\label{sec_conclusion}
We conclude by discussing directions of future research. The current paper takes the first step towards understanding  the  accountability and selection effects of personalized information aggregation.  We examine an otherwise classical accountability model for now,  and  hope to extend the analysis to alternative settings  featuring, e.g., bribery, corruption, and pandering, in the future.

In the current model, one wouldn't want to attend to the information acquired by the other voters even if he can, because the latter is a garbled signal of the underlying state and so is an inferior source of information compared to the original source. To qualify voters as secondary sources of information, we must enrich the current setting by, e.g., allowing for the possibility that attending to the other voters is more engaging  or less costly  than attending to the original  data source. Such an enrichment is shown to generate  echo chambers by \cite{network} in a model of multi-agent decision-making without political candidates; its implication  for electoral accountability and selection awaits future investigations.

As for how to test our theory, we believe that an important first step is to test the model of posterior separability in a political context. Our comparative statics exercises delineate the rich effects of varying voters' preferences and attention capacities on the outcome of personalized information aggregation.\footnote{While some of our results are qualitatively similar to the ones obtained by the existing studies on filtering bias, others (e.g., the effect on partisan disagreement as we vary voters' bandwidths; the comparative statics of the  incentive power generated by individual voters' signals when the underlying states are rich), are to the best of our knowledge new to the literature. \label{footnote_nonmonotonicity} }  One way to test these predictions is to conduct lab experiments that vary subjects' preferences and bandwidths in controlled environments as in \cite{ambuehl2017offer} and \cite{novak}. An another potentially useful approach is to study the experiments conducted by news aggregator companies or the regulatory uncertainties they face,  as they might generate the needed exogenous variations for empirical research (e.g., providing extreme voters with balanced viewpoints, as advocated by Allsides.com, is mathematically equivalent to reducing their partisan preference parameter).

\appendix

\numberwithin{equation}{section}
\counterwithin{figure}{section}
\section{Proofs}\label{sec_proof} 
\subsection{Useful observations and their proofs}\label{sec_proof_observation}
\paragraph{Proof of Observation \ref{obs_omega}} 
Let $\alpha$ denote the probability that the incumbent has ability $h$. Recall our normalization that 
\[
\mathbb{E}[\theta]=\alpha h+(1-\alpha)l=0.
\]
Since $\theta$ is a mean-preserving spread of $\mathbb{E}[\theta \mid \omega, a_h=1]$, 
the following inequality must hold:  $h>1>-1>l$.

In Footnote \ref{footnote_assumption}, we demonstrated that Assumption \ref{assm_singledimensional}(i): $\omega=\mathbb{E}[\theta\mid \omega, a_h=1]$ $\forall \omega \in \Omega$, is w.l.o.g.  Meanwhile, the restrictions imposed by Assumption \ref{assm_singledimensional}(ii): $\Omega=\{-1,1\}$,  on model primitives are threefold: 
\[
\begin{cases}
\mathbb{E}[\theta \mid \omega=1, a_h=1]=\frac{\alpha h p_1(1)+(1-\alpha)lp_0(1)}{\alpha p_1(1)+(1-\alpha)p_0(1)}=1,\\
\mathbb{E}[\theta \mid \omega=-1, a_h=1]=\frac{\alpha h p_1(-1)+(1-\alpha)lp_0(-1)}{\alpha p_1(-1)+(1-\alpha)p_0(-1)}=-1,\\
\mathbb{E}[\mathbb{E}[\theta \mid \omega, a_h=1] \mid a_h=1]=\mathbb{E}[\theta]=0.
\end{cases}
\]
The first two conditions require that the random variable $\mathbb{E}[\theta\mid \omega, a_h=1]$ takes values $\pm 1$. The last condition is nothing but Bayes' rule. Simplifying the first two conditions yields 
\begin{equation}\label{eqn_ll1}
\frac{p_1(1)}{p_0(1)}=\frac{(1-\alpha)(|l|+1)}{\alpha (h-1)}
\end{equation}
and 
\begin{equation}\label{eqn_ll2}
\frac{p_1(-1)}{p_0(-1)}=\frac{(1-\alpha)(|l|-1)}{\alpha (h+1)}. 
\end{equation}
Since (\ref{eqn_ll1})$>$(\ref{eqn_ll2}), Part (i) of Observation \ref{obs_omega} must hold. 

To simplify the last condition, write $\overline{p}(\omega)$ for the probability $\alpha p_1(\omega)+(1-\alpha)p_0(\omega)$ that the performance state is $\omega$ in case $a_h=1$.  Then the last condition is  simply  $\overline{p}(1)-\overline{p}(-1)=2\overline{p}(1)-1=0$,  or equivalently
\begin{equation}\label{eqn_pbar}
\overline{p}(1)=\frac{1}{2}
\end{equation}
as in Part (ii) of Observation \ref{obs_omega}.

Conditions (\ref{eqn_ll1})-(\ref{eqn_pbar}) constitute all the restrictions that $\mathbb{E}[\theta]=0$ and $\Omega=\{-1,1\}$ together impose on the model primitives. Solving these conditions simultaneously yields infinitely many solutions. \qed

\paragraph{Further observations} Fix $a_h=1$, and let $\Pi: \Omega \rightarrow \Delta(\mathcal{Z})$ be any finite signal structure. Since $\Omega$ is binary, it is w.l.o.g. to identify $\Pi$ with the tuple $(\pi_z, \mu_z)_{z \in \mathcal{Z}}$, where $\pi_z$ is the probability that the signal realization is $z$, and $\mu_z$ is the posterior mean of $\omega$ given $z$. In Section \ref{sec_discussion}, we derived the expressions for $\pi_z$ and $\mu_z$ for any given $\Pi$. Conversely, we can back out $\Pi$ based on $(\pi_z, \mu_z)_{z \in \mathcal{Z}}$. In what follows, we will  work with $(\pi_z, \mu_z)_{z \in \mathcal{Z}}$ because it is easy to deal with. Bayes' plausibility mandates that the expected posterior mean of $\omega$ must equal the prior mean zero:
\begin{equation}\label{eqn_bp}
\tag{BP}\sum_{z \in \mathcal{Z}} \pi_z \cdot \mu_z=0.
\end{equation}

For any binary signal structure, we write $\mathcal{Z}=\left\{L,R\right\}$, and assume w.l.o.g. that $\mu_L<0<\mu_R$. Bayes' plausibility (\ref{eqn_bp}) implies that
\begin{equation}\label{eqn_pi}
 \pi_L=\frac{\mu_R}{|\mu_L|+\mu_R} \text{ and }\pi_R=\frac{|\mu_L|}{|\mu_L|+\mu_R},
\end{equation}
so it is w.l.o.g. to identify $(\pi_z, \mu_z)_{z \in \{L, R\}}$   with the profile $\left(|\mu_L|, \mu_R\right)$ of posterior means (hereinafter written as $\left(x,y\right)$). From consuming $\left(x,y\right)$, voter $k$'s gains the following amount of expressive voting utility: 
\begin{equation}\label{eqn_v}
V_k\left(x,y\right)=\begin{cases}
\frac{x}{x+y}\left[v_k+y\right]^+ & \text{ if } k \leq 0,\\
-\frac{y}{x+y}\left[v_k-x\right]^- & \text{ if } k>0,
\end{cases}
\end{equation}
and incurs the following amount of attention cost: 
\begin{equation}\label{eqn_attentionbinary}
I\left(x,y\right)=\frac{y}{x+y}h\left(x\right)+\frac{x}{x+y}h\left(y\right). 
\end{equation}
A typical level curve of the attention cost function is thus
\begin{equation}\label{eqn_levelcurve}
\mathcal{C}\left(I\right)=
\left\{\left(x,y\right): \frac{y}{x+y} h\left(x\right)+\frac{x}{x+y}h\left(y\right)=I\right\},
\end{equation}
which is downward sloping by Assumption \ref{assm_attention}. 
Among all the signal structures lying on $\mathcal{C}\left(I\right)$, only $\left(h^{-1}\left(I\right), h^{-1}\left(I\right)\right)$ is neutral, whereas the remaining signal structures are either $L$-biased ($x<y$), or $R$-biased ($x>y$). For any $\left(x,y\right)$, $\left(x',y'\right) \in \mathcal{C}\left(I\right)$, either $\left(x,y\right)$ is more $L$-biased than $\left(x',y'\right)$ ($x/y<x'/y'$), or $(x,y)$ is more $R$-biased than $\left(x', y'\right)$ ($x/y>x'/y'$).

\subsection{Proofs of lemmas}\label{sec_proof_lemma}

\paragraph{Proof of Lemma \ref{lem_binary}} W.l.o.g. consider the problem faced by voter $1$. Any optimal personalized signal for him must solve
\begin{equation}\label{eqn_problem}
 \max_{\Pi: \Omega \rightarrow \Delta\left(\mathcal{Z}\right)} V_1\left(\Pi\right)
 \text{ s.t. }  I\left(\Pi\right)\leq I_1. 
\end{equation}
Let $\lambda \geq 0$ denote a Lagrange multiplier associated with the bandwidth constraint, and define the Lagrangian function as
\begin{align*}
    \mathcal{L}(\Pi,\lambda)=V_1\left(\Pi\right)+\lambda (I_1- I(\Pi)). 
\end{align*}
Then the primal problem: (\ref{eqn_problem}), can be rewritten as $\sup_{\Pi} \inf_{\lambda \geq 0} \mathcal{L}(\Pi,\lambda)$, whereas the dual problem is  $\inf_{\lambda \geq 0} \sup_{\Pi} \mathcal{L}(\Pi,\lambda)$. Let $p^*$ and $d^{*}$ denote the values of the primal problem and dual problem, respectively. Note that $p^*>0$ because $I_1>0$, and that $d^* \geq p^*$ by weak duality. 

The remainder of the proof consists of three steps. 

\paragraph{Step 1.}  Characterize the solution to $\sup_{\Pi} \mathcal{L}(\Pi,\lambda)$ for each $\lambda \geq 0$. Show that the solution is unique and has at most two signal realizations. 

When $\lambda=0$, the solution to $\sup_{\Pi} \mathcal{L}(\Pi,\lambda)$ is to fully reveal the state to the voter. When $\lambda>0$, rewrite the problem as 
\begin{equation}\label{eqn_relaxed}
\sup_{\Pi} V_1(\Pi)-\lambda I(\Pi), 
\end{equation}
or equivalently 
\[
   \sup_{\Pi} \sum_{z \in \mathcal{Z}} \pi_z \underbrace{\left(-\left[v_1+\mu_z\right]^{-}-\lambda h\left(\mu_z\right)\right)}_{f\left(\mu_z\right)}, 
\]
where $f$ is the maximum of two strictly concave functions of $\mu$:  $-\lambda h\left(\mu\right)$ and $ -v_1-\mu-\lambda h\left(\mu\right)$. Since these functions single-cross at  $\mu=-v_1$ and their maximum is M-shaped, solving \eqref{eqn_relaxed} using the concavification method developed by \cite{bayesianpersuasion} yields a unique solution with at most two signal realizations. Denote this solution by  $\Pi(\lambda)$, and note that it is continuous in $\lambda$ by Berge's maximum theorem.

\paragraph{Step 2.} Show that strong duality holds, i.e., $p^*=d^*$, and that the value is attained at a positive, finite, $\lambda^*$. 

Since $d^* \geq p^*>0$ and $d^* \leq \mathcal{L}(\Pi(0), 0)=V_1(\Pi(0))$, $d^*$ must be a positive, finite number. By the envelope theorem, $\mathcal{L}(\Pi(\lambda), \lambda)$ is absolutely continuous in $\lambda$ (i.e., differentiable in $\lambda$ almost surely), and the derivative, whenever it exists, is given by  
\[
    \frac{d}{d \lambda} \mathcal{L}(\Pi(\lambda),\lambda)= I_1 -I(\Pi(\lambda)).
\]
Now, since the right-hand side of the above expression is continuous in $\lambda$ (as shown in Step 1), $\mathcal{L}(\Pi(\lambda),\lambda)$ must be differentiable in $\lambda$ rather than being just absolutely continuous in $\lambda$ on $(0, +\infty)$. Let $\lambda^*$ be a Lagrange multiplier that attains $d^*$. Note first that $\lambda^* \neq 0$, because at $\lambda=0$, $\Pi(0)$ reveals the true state to the voter and, under Assumption \ref{assm_attention}(iii),  satisfies $\Pi(0)>I_1$. This means that $\mathcal{L}(\Pi(\lambda),\lambda)$ is decreasing in $\lambda$ on $\in [0, \epsilon)$ for some $\epsilon>0$, hence $\mathcal{L}$ isn't minimized at $\lambda=0$. 

There are two remaining possibilities to consider. 

\subparagraph{Case 1.} $\lambda^* \in (0, +\infty)$. In this case, we must have 
\[
    \frac{d}{d \lambda} \mathcal{L}(\Pi(\lambda),\lambda))\bigg|_{\lambda=\lambda^*} = I_1 -I(\Pi(\lambda^*))=0,
\]
hence $(\lambda^*, \Pi(\lambda^*))$ satisfies the complementary slackness condition and so is primal feasible. This implies that $p^* \geq \mathcal{L}(\Pi(\lambda^*), \lambda^*)=d^* \geq p^*$ as desired.

\subparagraph{Case 2.} $\lambda^*=+\infty$.  In this case $\lim_{\lambda \rightarrow +\infty} I(\Pi(\lambda))=I_1$ must hold in order to prevent $\mathcal{L}$ from exploding in the limit.  Meanwhile, $\Pi(+\infty)$ solves $\max_{\Pi} V(\Pi)-+\infty \cdot I(\Pi)$ and so must be degenerate, which together with the continuity of $\Pi(\lambda)$ in $\lambda$ implies that  $\Pi(\lambda)<I_1$ when $\lambda$ is sufficiently large, a contradiction. 

Taken together, we conclude that strong duality holds, and that $p^*=d^*$ is attained at a positive, finite $\lambda^*$.  $\Pi(\lambda^*)$ is a binary signal structure and satisfies $I(\Pi(\lambda^*))=I_1$. 

\paragraph{Step 3.} Show that the primal problem admits a unique solution that satisfies all properties stated in the lemma. From the previous steps, we know that if the primal problem admits two distinct solutions, then they must take the form of $\Pi(\lambda^1)$ and $\Pi(\lambda^2)$ for some $\lambda^1, \lambda^2>0$ such that $\lambda^1 \neq \lambda^2$. Assume w.l.o.g. that $\lambda^1>\lambda^2$. Since $\Pi(\lambda)$ is the unique solution to $V_1(\Pi)-\lambda I(\Pi)$, the following inequalities must hold: 
\[
\lambda^1 \left(I\left(\Pi^2 \right)-I\left( \Pi^1 \right)\right) > V_1\left( \Pi^2 \right)-V_1\left( \Pi^1 \right) > \lambda^2 \left(I\left(\Pi^2 \right)-I\left(\Pi^1 \right)\right).
\]
Then from $\lambda^1 >\lambda^2$, it follows that $I\left(\lambda^2 \right)>I\left(\lambda^1 \right)$, which contradicts the fact that $I\left(\Pi^1\right)=I\left(\Pi^2\right)=I_1$.  

It remains to show that the posterior means of the state induced by $\Pi(\lambda^*)$ satisfies $v_1+\mu_L<0<v_1+\mu_R$. Since $\mu_L<0<\mu_R$, $v_1+\mu_R>0$ holds automatically. To see why $v_1+\mu_L<0$, note that if the contrary is true, i.e., $v_1+\mu_L \geq 0$, then  $V_1(\Pi)=-\pi_L[v_1+\mu_L]^-=0$, hence the voter is strictly better-off by acquiring no information and voting unconditionally for his own-party candidate $R$, a contradiction.  \qed

\paragraph{Proof of Lemma \ref{lem_skewness}} 
Again, we only prove the result for voter $1$, and in two steps. 

\paragraph{Step 1.} Show that $\Pi_1$ is $R$-biased. Write $\left(x,y\right)$ for $\Pi_1$. Notice first that $\left(x,y\right)$ cannot be $L$-biased, because if the contrary were true, i.e., $y>x$, then voter $1$ would strictly prefer $\left(y,x\right)$ to $\left(x,y\right)$:
\[V_1\left(y,x\right)=\frac{x}{x+y}\left(y-v_1\right)>\frac{y}{x+y}\left(x-v_1\right)=V_1\left(x,y\right),\]
and yet $(x,y)$ and $(y,x)$ incur the same attention cost by Assumption \ref{assm_attention}. It remains to show that $\left(x,y\right)$ isn't neutral, i.e., $x\neq y$. For starters, rewrite (\ref{eqn_relaxed}) as 
\begin{equation}\label{eqn_relaxed1}
\max_{x \in \left[v_1,1\right], y \in \left[0,1\right]}  \text{ }  \frac{y}{x+y}\left(x-v_1\right)-\lambda\left(\frac{y}{x+y} h\left(x\right)+\frac{x}{x+y}h\left(y\right)\right).
\end{equation}
If the solution to (\ref{eqn_relaxed1}) were neutral, i.e., $x=y$, then only three situations can happen: $\left(x,y\right) \in \left(v_1,1\right)^2$, $\left(x,y\right)=\left(1,1\right)$, and $\left(x,y\right)=\left(v_1,v_1\right)$. In the first situation, $\left(x,y\right)$ must satisfy the following first-order conditions: 
\begin{align}
y+v_1 &=\lambda\left(\Delta  +h'\left(x\right)\Sigma\right)\label{eqn_foc1} \\
\text{ and } x-v_1 &=\lambda\left(h'\left(y\right)\Sigma-\Delta \right)\label{eqn_foc2}
\end{align}
where $\Delta \coloneqq h\left(y\right)-h\left(x\right)$ and $\Sigma\coloneqq x+y$. Plugging $x=y$ into (\ref{eqn_foc1}) and (\ref{eqn_foc2}) and simplifying yields $x+v_1=\lambda h'\left(x\right) \cdot 2x=x-v_1$, which is impossible. Meanwhile, the second situation is impossible because the voter would run out of bandwidth.  In the third situation, we have $V_1\left(v_1,v_1\right)=0$.  But then the voter would strictly prefer $\left(1-\epsilon, \epsilon\right)$ to $\left(v_1,v_1\right)$ when $\epsilon>0$ is sufficiently small, because the former generates a strictly positive expressive voting utility gain: 
\[V_1\left(1-\epsilon, \epsilon\right)=\frac{\epsilon}{1-\epsilon+\epsilon}\left(1-\epsilon-v_1\right)>0, \]
and it is moreover feasible: 
\[I\left(1-\epsilon, \epsilon\right)=\frac{\epsilon}{1-\epsilon+\epsilon} h\left(1-\epsilon\right)+\frac{1-\epsilon}{1-\epsilon+\epsilon}h\left(\epsilon\right)<\epsilon h\left(1\right)+h\left(\epsilon\right)<I_1.\]

\paragraph{Step 2.} Show that $\Pi_1$ becomes more $R$-biased as $v_1$ increases. Fix any $0<v_1<v_1'$, and let $\left(x,y\right)$ and $\left(x',y'\right)$ denote the unique solutions to (\ref{eqn_problem}) when the voter's partisan preference parameters are given by $v_1$ and $v_1'$, respectively. From the fact that the voter strictly prefers $\left(x,y\right)$ to $\left(x',y'\right)$ when his partisan preference parameter is $v_1$, and his preference is reversed when his partisan preference parameter is $v'_1$, i.e., 
\[\frac{y}{x+y}\left(x-v_1\right)>\frac{y'}{x'+y'}\left(x'-v_1\right) \text{ and } \frac{y'}{x'+y'}\left(x'-v_1'\right)>\frac{y}{x+y}\left(x-v_1'\right),\]
it follows that 
\[\left(\frac{y'}{x'+y'}-\frac{y}{x+y}\right)v_1>\frac{x'y'}{x'+y'}-\frac{xy}{x+y}>\left(\frac{y'}{x'+y'}-\frac{y}{x+y}\right)v_1'\]
and hence that $y'/\left(x'+y'\right)<y/\left(x+y\right)$. The last condition can be rewritten as $x'/y'>x/y$, which proves that $\left(x',y'\right)$ is more $R$-biased than $\left(x,y\right)$. \qed

\paragraph{Proof of Lemma \ref{lem_variable}} Let $\left(x_k, y_k\right)$ denote the optimal personalized signal for voter $k \in \mathcal{K}$. Tedious algebra (as detailed in the proof of Theorem \ref{thm_accountability})  shows that 
\begin{equation}\label{eqn_pd}
P_k=\frac{2x_k y_k}{x_k+y_k} \text{ } \forall k \in \mathcal{K} \text{ and } D=1-\frac{2x_1y_1\left(1+x_1 y_1 \right)}{\left(x_1+y_1\right)^2}.
\end{equation}
Write $P_1$ and $D$ as functions of $\left(x_1, y_1\right)$, or simply $\left(x,y\right)$.  For each function $g \in \left\{I, P_1, D \right\}$ of $\left(x,y\right)$, write $g_x$ for $\partial g\left(x,y\right)/\partial x$ and $g_y$ for $\partial g\left(x,y\right)/\partial y$. Then 
\begin{align}
&I_x =\frac{y}{\left(x+y\right)^2}\left[h\left(y\right)-h\left(x\right)+h'\left(x\right)\left(x+y\right)\right] \label{eqn_Ix}\\
&I_y=\frac{x}{\left(x+y\right)^2}\left[h\left(x\right)-h\left(y\right)+h'\left(y\right)\left(x+y\right)\right] \label{eqn_Iy}\\
&P_{1,x}=\frac{2y^2}{\left(x+y\right)^2} \label{eqn_Px} \\
&P_{1,y}=\frac{2x^2}{\left(x+y\right)^2} \label{eqn_Py}\\
&D_x=-\frac{2y\left(2xy^2-x+y\right)}{\left(x+y\right)^3} \label{eqn_Dx}\\
 \text{ and } &D_y=-\frac{2x\left(2x^2y-y+x\right)}{\left(x+y\right)^3}. \label{eqn_Dy}
\end{align}
Simplifying (\ref{eqn_Ix}) and (\ref{eqn_Iy}) using the assumption that $h$ is strictly increasing and strictly convex on $[0,1]$ yields $I_x, I_y>0$. Specifically,  $h(y)-h(x)+h'(x)(x+y)\geq h'(x)(y-x)+h'(x)(x+y)=2h'(x)y>0$, and $h(x)-h(y)+h'(y)(x+y)>0$ for the same reason.
 
\bigskip

\noindent Part (i): 
We first show that $P_1\left(x,y\right)$ decreases as we traverse along voter $1$'s attention level curve $\mathcal{C}\left(I_1\right)$ from its neutral element to its most $R$-biased element. This portion of the level curve can be expressed as 
\begin{equation}
\mathcal{C}^+\left(I_1\right)=\left\{\left(x,y\right) \in \mathcal{C}\left(I_1\right): x \in \left[h^{-1}\left(I_1\right), 1\right]\right\},
\end{equation}
and it satisfies (i) $x \geq y$, and (ii) $x>y$ if and only if $x>h^{-1}\left(I_1\right)$. As we increase $x$ by a small amount $\epsilon>0$, we must change  $y$ by approximately $\left(-I_x /I_y\right)\cdot \epsilon$ in order to stay on $\mathcal{C}^+\left(I_1\right)$. The resulting change in $P_1$ equals approximately $\left(P_{1,x}-P_{1,y} I_x/I_y\right)\cdot \epsilon$. Since $P_{1,y}>0$, it suffices to show that $-I_x/I_y<-P_{1,x}/P_{1,y}$ holds for all $\left(x,y\right) \in \mathcal{C}^+\left(I_1\right)$. Simplifying the last condition using (\ref{eqn_Ix})-(\ref{eqn_Py}) yields
\begin{equation}\label{eqn_skewness1}
\frac{h\left(y\right)-h\left(x\right)+h'\left(x\right)\left(x+y\right)}{h\left(x\right)-h\left(y\right)+h'\left(y\right)\left(x+y\right)}>\frac{y}{x}.
\end{equation}
To demonstrate the validity of (\ref{eqn_skewness1}), note that we can bound the numerator and denominator on its left-hand side as follows, using the assumption that $h$ is strictly convex and  strictly increasing on $\left[0,1\right]$: 
\begin{align*}
h\left(y\right)-h\left(x\right)+h'\left(x\right)\left(x+y\right)&>h'\left(x\right)\left(y-x\right)+h'\left(x\right)\left(x+y\right)=2h'\left(x\right)y,\\
h\left(x\right)-h\left(y\right)+h'\left(y\right)\left(x+y\right)&<h'\left(x\right)\left(x-y\right)+h'\left(x\right)\left(x+y\right)=2h'\left(x\right)x.
\end{align*}
Combining the above inequalities gives the desired result.

We next show that $D\left(x,y\right)$ increases as we traverse along $\mathcal{C}^+\left(I_1\right)$ as above. Note first that for any $\left(x,y\right) \in \mathcal{C}^+\left(I_1\right)$, $D_y=-2x\left(2x^2y-y+x\right)/\left(x+y\right)^3<0$ because $x\geq y$. Thus if $D_x\geq 0$, then $D_x-D_y I_x/I_y>0$, and we are done. If $D_x<0$, then it suffices to show that $-I_x/I_y<-D_x/D_y$, or equivalently
\[
\frac{h\left(y\right)-h\left(x\right)+h'\left(x\right)\left(x+y\right)}{h\left(x\right)-h\left(y\right)+h'\left(y\right)\left(x+y\right)}>\frac{2xy^2-x+y}{2x^2y-y+x}.
\]
The last condition follows from (\ref{eqn_skewness1}) and $x>y$, which together imply that 
\[\frac{h\left(y\right)-h\left(x\right)+h'\left(x\right)\left(x+y\right)}{h\left(x\right)-h\left(y\right)+h'\left(y\right)\left(x+y\right)}>\frac{y}{x}=\frac{2xy^2}{2x^2y}>\frac{2xy^2-x+y}{2x^2y-y+x}.\]

\bigskip

\noindent Part (ii): We first prove the claim that $P_k$ is increasing in $I_k$ for voter $1$. The proofs for voters $0$ and $-1$ are analogous and hence are omitted. Since $P_1\left(x,y\right)= 2xy/\left(x+y\right)$ is increasing in $x$ and $y$, it suffices to show that $x$ and $y$ are both increasing in $I_1$. The remainder of the proof proceeds in two steps.

\paragraph{Step 1.}  Show that $x$ and $y$ increase as the Lagrange multiplier associated with voter $1$'s bandwidth constraint decreases. Recall (\ref{eqn_relaxed1}), which says maximizing voter $1$'s expressive voting utility while taking the Lagrange multiplier $\lambda>0$ associated with his bandwidth constraint as given. If the solution to that  problem lies in the interior of $\left[v_1,1\right]\times \left[0,1\right]$, then it must satisfy the first-order conditions (\ref{eqn_foc1}) and (\ref{eqn_foc2}). Summing up (\ref{eqn_foc1}) and (\ref{eqn_foc2}) yields
\begin{equation}\label{eqn_sumfoc}
h'\left(x\right)+h'\left(y\right)=1/\lambda.
\end{equation}
Using this result when differentiating (\ref{eqn_foc2}) with respect to $\lambda$ yields 
\begin{align*}
-\frac{dx}{d\lambda}=&\Delta-h'\left(y\right)\Sigma\\
&+\lambda\left[h'\left(y\right)\frac{dy}{d\lambda}-h'\left(x\right)\frac{dx}{d\lambda}-h''\left(y\right)\frac{dy}{d\lambda}\Sigma - h'\left(y\right)\frac{dy}{d\lambda}-h'\left(y\right)\frac{dx}{d\lambda}\right]\\
=&\Delta -h'\left(y\right)\Sigma-\lambda h''\left(y\right)\frac{dy}{d\lambda}\Sigma-\frac{dx}{d\lambda}
\end{align*} 
where $\Delta \coloneqq h\left(y\right)-h\left(x\right)$ and $\Sigma\coloneqq x+y$.
Therefore, 
\begin{equation}\label{eqn_dydlambda}
\frac{dy}{d\lambda}=\frac{\Delta-h'\left(y\right)\Sigma}{\lambda h''\left(y\right)\Sigma}=\frac{h\left(y\right)-h\left(x\right)-h'\left(y\right)\left(x+y\right)}{\lambda h''\left(y\right)\left(x+y\right)}<0,
\end{equation}
where the last inequality exploits the assumption that $h''>0$, $h'>0$ on $\left(0,1\right)$ (hence $h\left(y\right)-h\left(x\right)-h'\left(y\right)\left(x+y\right)<0$ if $0\leq y<x$ and $h\left(y\right)-h\left(x\right)-h'\left(y\right)\left(x+y\right)\leq h'\left(y\right)(y-x)-h'\left(y\right)\left(x+y\right)<0$ if $y\geq x (\geq v_1)$). Meanwhile, differentiating (\ref{eqn_sumfoc}) with respect to $\lambda$ yields
\[-h''\left(x\right)\frac{dx}{d\lambda}=h''\left(y\right)\frac{dy}{d\lambda}+\frac{1}{\lambda^2}.\]
Simplifying this result using (\ref{eqn_sumfoc}) and (\ref{eqn_dydlambda}) yields
\begin{equation}\label{eqn_dxdlambda}
\frac{dx}{d\lambda}=-\frac{\Delta+h'\left(x\right)\Sigma}{\lambda h''\left(x\right) \Sigma}=-\frac{h\left(y\right)-h\left(x\right)+h'\left(x\right)\left(x+y\right)}{\lambda h''\left(x\right) \left(x+y\right)}<0,
\end{equation}
where the last inequality follows again from the assumption that $h''>0$, $h'>0$ on $(0,1)$ (hence $h\left(y\right)-h\left(x\right)+h'\left(x\right)(x+y)>0$ if $y\geq x (\geq v_1)$ and $h\left(y\right)-h\left(x\right)+h'\left(x\right)(x+y)>-h'(x)(x-y)+h'\left(x\right)(x+y)>0$ if $0\leq y<x$). Together, (\ref{eqn_dydlambda}) and (\ref{eqn_dxdlambda}) imply that $x$ and $y$ strictly increase as $\lambda$ slightly decreases. As $\lambda$ further decreases, the solution to (\ref{eqn_relaxed1}) may transition from an interior one to a corner one. When that happens, we must have $x=1$, because $\left(x,y\right)$ is $R$-biased by Lemma \ref{lem_skewness}. As $\lambda$ continues to decrease, $x$ stays at $1$ whereas $y$ increases. 

\paragraph{Step 2.} Show that the Lagrange multiplier associated with voter $1$'s bandwidth constraint decreases with his bandwidth. Take any $0<I<I'$. Let $\lambda$ and $\lambda'$ denote the Lagrange multipliers associated with voter $1$'s bandwidth constraint when his bandwidths are given by $I$ and $I'$, respectively, and let $\left(x,y\right)$ and $\left(x',y'\right)$ denote the solutions to (\ref{eqn_relaxed1}) given $\lambda$ and $\lambda'$, respectively. From strict optimality, i.e., voter $1$ strictly prefers $\left(x,y\right)$ to $\left(x',y'\right)$ when the Lagrange multiplier is $\lambda$, and his preference is reversed when the Lagrange multiplier is $\lambda'$, we deduce that 
\[\lambda\left[I\left(x',y'\right)-I\left(x,y\right)\right]>V_1\left(x',y'\right)-V_1\left(x,y\right)>\lambda'\left[I\left(x',y'\right)-I\left(x,y\right)\right].\]
Then from $I\left(x,y\right)=I<I'=I\left(x',y'\right)$, it follows that $\lambda'<\lambda$, which together with the result shown in Step 1 implies that  $x' \geq x$, $y' \geq y$, and one of these inequalities is strict.  

\bigskip

We next show that $D$ is in general non-monotonic in voter $1$'s bandwidth or,  equivalently, the Lagrange multiplier $\lambda>0$ associated with his bandwidth constraint. The total derivative of $D$ w.r.t. $\lambda$ equals  
$
D_x dx/d\lambda + D_y dy/d\lambda, 
$
which is positive if and only if $D_x/|D_y|<|dy/d\lambda|/|dx/d\lambda|$. The last condition is automatically satisfied if $D_x<0$. If $D_x>0$, then we can deduce, from (\ref{eqn_Dx}) and (\ref{eqn_Dy}), that $x>y$, and rewrite the last condition as   
\begin{equation}\label{eqn_Dlambda}
\underbrace{-\frac{y\left(2xy^2-x+y\right)}{x\left(2x^2y-y+x\right)}}_{\in (0,1)}<\frac{h\left(x\right)-h\left(y\right)+h'\left(y\right)\left(x+y\right)}{h\left(y\right)-h\left(x\right)+h'\left(x\right)\left(x+y\right)}\cdot \frac{h''\left(x\right)}{h''\left(y\right)}.
\end{equation} 
Condition (\ref{eqn_Dlambda}) may or may not hold for functions that satisfy Assumption \ref{assm_attention}. It holds for $h_1(x)=x^2$, which, when plugged into the right-hand side of (\ref{eqn_Dlambda}), sets the result equal to one. To construct a function that violates (\ref{eqn_Dlambda}), 
fix any $x^* \in \left(0,1\right)$ and any $\epsilon>0$ that is arbitrarily small. Let $\alpha, \beta$ be such that $h_2(x)\coloneqq \alpha x^{1+\epsilon}+\beta$ satisfies $h_1(x^*+\epsilon)=h_2(x^*+\epsilon)$ and $h_1'(x^*+\epsilon)=h_2'(x^*+\epsilon)$. Based on $h_1$ and $h_2$, define a new function $h: \left[-1,1\right]\rightarrow \mathbb{R}_+$ where $h(x)=h(-x)$, $h\left(x\right)=h_1(x)$ if $x \in \left[0,x^*\right]$, and $h\left(x\right)=h_2(x)$ if $x \in \left[x^*+\epsilon, 1\right]$. Over $\left[x^*, x^*+\epsilon\right]$, let $h$ be any function that satisfies $h'>0$, $h''>0$,  and smooth-pasting at $x=x^*$ and $x^*+\epsilon$,  and we are done. \qed 

\subsection{Proofs of theorems and propositions}\label{sec_proof_theorem}
\paragraph{Proof of Theorem \ref{thm_accountability}} Write $a_k^+$ for $\Pi_k\left(R \mid \omega=1\right)$ and $a_k^-$ for $\Pi_k\left(R \mid \omega=-1\right)$. Symmetry implies that \[a_{-1}^+=1-a_1^-,  \text{ } a_{1}^+=1-a_{-1}^-, \text{ and } a_0^+=1-a_0^-.\] Write $\left(x_k, y_k\right)$ for $\Pi_k$, and note that \[a_k^+=\frac{x_k\left(1+y_k\right)}{x_{k}+y_k} \text{ and }a_k^-=\frac{x_k\left(1-y_k\right)}{x_{k}+y_k}.\] 
Thus
\[
P_k \coloneqq a_k^+-a_k^-=\frac{2x_ky_k}{x_k+y_k} \text{ } \forall k \in \mathcal{K}\]
\[\text{ and } D \coloneqq \mathbb{P}_{\Pi_ks}\left(z_{-1}\neq z_1 \mid a_h=1\right)
=\frac{1}{2}\sum_{\omega\in \Omega} \mathbb{P}_{\Pi_ks}\left(z_{-1}\neq z_1\mid \omega\right)
=1-\frac{2x_1y_1\left(1+x_1y_1\right)}{(x_1+y_1)^2}, 
\]
where the last equality can be established as follows: 
\begin{align*}
&\mathbb{P}_{\Pi_ks}\left(z_{-1}\neq z_1 \mid \omega=1\right)\\
&=a_{-1}^+\left(1-a_1^+\right)+ \left(1-a_{-1}^+\right)a_1^+\\
\tag{$\because $ symmetry}&=\left(1-a_1^-\right)a_{-1}^-+ a_1^-\left(1-a_{-1}^-\right)\\
&=\mathbb{P}_{\Pi_ks}\left(z_{-1}\neq z_1 \mid \omega=-1\right)\\
\tag{$\because $ symmetry}&=\left(1-a_1^-\right)\left(1-a_1^+\right)+a_1^-a_1^+\\
&=1-\frac{2x_1y_1\left(1+x_1y_1\right)}{(x_1+y_1)^2}. 
\end{align*}
The remainder of the proof proceeds in three steps.

\paragraph{Step 1.}  Reduce $\xi$ to model primitives. Recall that 
\[\xi\coloneqq \mathbb{P}_{\Pi_ks}\left(\text{R wins re-election}\mid \omega=1\right)-\mathbb{P}_{\Pi_ks}\left(\text{R wins re-election}\mid \omega=-1\right).\] Consider two cases:  $f_0 \geq 1/2$ and $f_0<1/2$. In the first case, candidate $R$ wins re-election if and only if voter $0$'s signal recommends $R$, so $\xi=P_0$. 
In the second case, candidate $R$ wins re-election in two events: (i) voter $\pm 1$'s signals both recommend $R$; (ii) voter $\pm 1$'s signals send conflicting recommendations, and voter $0$'s signal recommends $R$. The part of $\xi$ that stems from event (i) equals 
\[
a_{-1}^{+}a_1^{+}-a_{-1}^{-}a_1^{-}=\left(1-a_1^-\right)a_1^{+}-\left(1-a_1^+\right)a_1^{-}=a_1^{+}-a_1^{-}\coloneqq P_1. 
\]
The part of $\xi$ that stems from event (ii) equals 
\begin{align*}
&\mathbb{P}_{\Pi_ks}\left(z_{-1}\neq z_1, z_0=R\mid \omega=1\right)-\mathbb{P}_{\Pi_ks}\left(z_{-1}\neq z_1, z_0=R \mid \omega=-1\right)\\
&=Da_0^{+}-Da_0^-=DP_0.
\end{align*}
Summing things up yields $\xi=P_1+DP_0$.

\paragraph{Step 2.} Analyze when accountability can arise in an equilibrium.  In  case voters use $\Pi_k$s, candidate $R$ changes his re-election probability by the following amount as he raises his effort level from low to high: 
\begin{align*}
&\sum_{\omega \in \Omega}p_{1}\left(\omega\right) \mathbb{P}_{\Pi_k s}\left(R \text{ wins re-election} \mid \omega\right)-p_{0}\left(\omega\right) \mathbb{P}_{\Pi_k s}\left(R \text{ wins re-election} \mid \omega\right)\\
&=\left(p_1\left(1\right)-p_0\left(1\right)\right)\xi.
\end{align*}
Exerting high effort is a best response to $\Pi_k$s if and only if $\left(p_1\left(1\right)-p_0\left(1\right)\right)\xi \geq c$ or, equivalently, $\xi \geq \hat{c}$. 

\paragraph{Step 3. } Solve for the degree of electoral selection. Without accountability, performance data carry no information about the incumbent's ability, which makes  selection impossible. With accountability, the degree of electoral selection equals 
\begin{multline*}
\frac{1}{2} \sum_{\omega \in \Omega}\mathbb{P}_{\Pi_ks}\left(R \text{ wins re-election} \mid \omega\right) \cdot \omega + \mathbb{P}_{\Pi_ks}\left(L \text{ wins election} \mid \omega\right) \cdot 0\\
=\frac{1}{2} \left[\mathbb{P}_{\Pi_ks}\left(R \text{ wins re-election} \mid \omega=1\right)-\mathbb{P}_{\Pi_ks}\left(R \text{ wins re-election} \mid \omega=-1\right)\right]
=\frac{\xi}{2}.\qed 
\end{multline*}

\paragraph{Proof of Proposition \ref{prop_skewness}} When $P_0=0$, $\xi=P_1+D \cdot 0=P_1$ and so is decreasing in $v_1$ by Lemma \ref{lem_variable}. When $P_0=1$, 
\[
\xi=P_1+D\cdot 1=1-\frac{2x_1 y_1\left(1-x_1\right)\left(1-y_1\right)}{\left(x_1+y_1\right)^2 },
\]
where the last expression is maximized when $x_1=1$ (equivalently, when $\left(x_1, y_1\right)$ is the most $R$-biased element of the level curve $\mathcal{C}\left(I_1\right)$ of the attention cost function).  For intermediate values of $P_0$, $\xi$ can vary non-monotonically with $v_1$ as demonstrated in  Example \ref{exm_monotone}. \qed

\paragraph{Proofs of Propositions \ref{prop_bandwidth} and \ref{prop_population}}  Results follow immediately from Lemma \ref{lem_variable} and \ref{thm_accountability}. \qed

\section{Continuous state distribution}\label{sec_state}
In this appendix, consider a variant of the baseline model where $\Omega=\left[-1,1\right]$ and the performance state generated by an effort choice $a \in \left\{0,1\right\}$ has a p.d.f.  $p_a$ that is positive almost everywhere. Let $\alpha$ denote the probability that the incumbent has ability $h$, and define $\overline{p}\coloneqq \alpha p_1+(1-\alpha)p_0$ as the p.d.f. of the performance state in case $a_h=1$. Suppose w.l.o.g. that $\omega=\mathbb{E}\left[\theta \mid \omega; a_h=1\right]$. 

A single voter with partisan preference parameter $-v \in\left(-1,0\right)$ and bandwidth $I >0$ can acquire any finite signal $\Pi: \Omega \rightarrow \Delta\left(\mathcal{Z}\right)$ about $\omega$. When  $a_h=1$,
\[\pi_z=\int \omega \Pi\left(z \mid \omega\right) \overline{p}\left(\omega\right)d\omega\]
is the probability that the signal realization is $z$, and 
\[\mu_z=\frac{\int \omega \Pi\left(z \mid \omega\right)\overline{p}\left(\omega\right)d\omega}{\int\Pi\left(z \mid \omega\right)\overline{p}\left(\omega\right)d\omega}\]
is the posterior mean of the performance state conditional on the signal realization being $z$. 
Voter's expected utility gain from consuming $\Pi$ equals
\[\sum_{z \in \mathcal{Z}}\pi_z[-v+\mu_z]^+,\]
and the attention cost associated with acquiring $
\Pi$ is 
\[I\left(\Pi\right)=H\left(\overline{p}\right)-\mathbb{E}_{\Pi}\left[H\left(\overline{p}\left(\cdot \mid \omega\right)\right)\right],\] 
where $H$ denotes the entropy function. To make the problem of information aggregation nontrivial, suppose that $H(\overline{p})>I$. Under this assumption,  the solution to the voter's problem (\ref{eqn_problem}) must satisfy Lemma \ref{lem_binary} by \cite{mckay}.  The incentive power $P$ generated by the solution is 
\[\int m(\omega)\left(p_1\left(\omega\right)-p_0\left(\omega\right)\right) d\omega,\]
where $m(\omega)\coloneqq \Pi\left(R \mid \omega\right)$. 

When $\omega$ is binary, $P$ is always decreasing in $v$ as shown in Lemma \ref{lem_variable}. With a continuum of states,  $P$ can increase rather than decrease with $v$ as demonstrated by the next example.
\begin{example}\label{exm_entropy}
Let $h=1$, $l=-1$, $\alpha=1/2$, and $p_1\left(\omega\right)=\left(1+\omega\right)/2$. From $\omega=\mathbb{E}[\theta \mid \omega, a_h=1]$, it follows that $p_0\left(\omega\right)/p_1\left(\omega\right)=\left(h-\omega\right)/\left(\omega-l\right)$ and, hence, that $p_0(\omega)=\left(1-\omega\right)/2$, $\overline{p}(\omega)=1/2$, and $P=\int \omega m\left(\omega\right) dw$. When $I=.1$, solving $P$ for $v=.24$ and $.25$ numerically yields $.13$ and $.14$, respectively. 

To see why $P$ can increase rather than decrease with $v$, note that when $v=.25$, the voter is biased towards candidate $L$ and so focuses his attention mainly on distinguishing whether $\omega$ is close to $1$ or not. The resulting function $m$ is flat and takes small values for most $\omega$s, but it rises sharply as $\omega$ approaches $1$ as depicted in Figure \ref{figure_mv}. 

\begin{figure}[!h]
\centering
     \includegraphics[width=.4\textwidth]{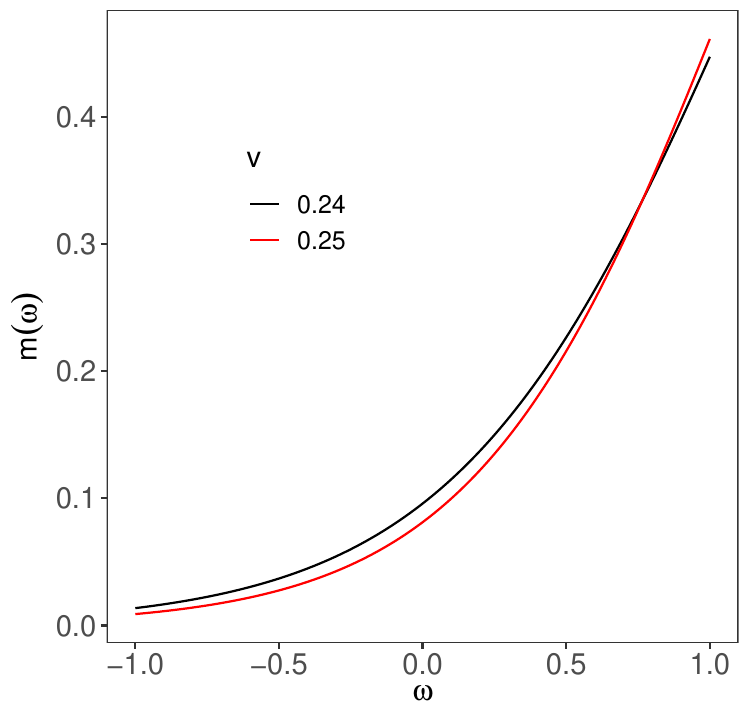}
    \caption{Plot $m\left(\omega\right)$ against $\omega$: model parameters are as specified in the text.}\label{figure_mv}
\end{figure} 
 As $v$ decreases from $.25$ to $.24$, the voter becomes more moderate and so allocates his attention more evenly across the various states.  Moreover,  his average propensity to vote for candidate $R$ increases, hence the function $m$ takes a higher value on average but is flatter around $\omega \approx 1$ than in the previous case.  
 
 To finalize the comparison, recall that $P=\int \omega m\left(\omega\right) d\omega$.  When the complementarity between $\omega$ and $m\left(\omega\right)$ is sufficiently strong around $\omega \approx 1$ in the first case (as in the current example),  raising $v$ could increase rather than decrease $P$ as claimed. $\hfill \diamondsuit$
\end{example}

 \section{Correlated signals}\label{sec_correlation}
In this appendix,  consider joint signal distributions whose marginal distributions are as solved in Section \ref{sec_analysis}. Let $a_1-a_8$ and $b_1-b_8$ denote the probabilities of receiving the various voting recommendation profile $z_{-1}z_0z_1$'s in state $\omega=1$ and $\omega=-1$, respectively:
\begin{center}
\begin{tabular}{l*{7}{c}r}
Prob.             & LLL & RLL & LRL & LLR & RRL  & LRR & RLR & RRR  \\
\hline
$\omega=1$ & $a_8$ & $a_7$ & $a_6$ & $a_5$ & $a_4$ & $a_3$ & $a_2$  &  $a_1$\\
 $\omega=-1$ & $b_8$ & $b_7$ & $b_6$ & $b_5$ & $b_4$ & $b_3$ & $b_2$  &  $b_1$ \\
\end{tabular}
\end{center}
These probabilities must satisfy \emph{feasibility}:
\begin{equation}\label{feasibility}
a_i, b_i \geq 0 \text{ } \forall i=1,\cdots, 8, \text{ } \sum_{i=1}^8a_i=1, \text{ and } \sum_{i=1}^8 b_i=1,
\end{equation}
as well as \emph{consistency}:
\begin{align}
& a_1+a_2+a_3+a_5 = \Pi_1\left(R \mid \omega=1\right) \nonumber \\
& a_1+a_3+a_4+a_6  = \Pi_0\left(R \mid \omega=1\right)\nonumber \\
& a_1+a_2+a_3+a_7  = \Pi_{-1}\left(R \mid \omega=1\right) \nonumber \\
& b_1+b_2+b_3+b_5  = \Pi_1\left(R \mid \omega=-1\right)\nonumber \\
& b_1+b_3+b_4+b_6  = \Pi_0\left(R \mid \omega=-1\right)\nonumber \\
\text{ and } & b_1+b_2+b_3+b_7  = \Pi_{-1}\left(R \mid \omega=-1\right).\label{consistency}
\end{align}
In the case where $f_0<1/2$, the societal incentive power generated by $\left\{a_i,b_i\right\}_{i=1}^8$ equals 
\begin{equation}\label{joint}
a_1+a_2+a_3+a_4-\left(b_1+b_2+b_3+b_4\right),
\end{equation}
Maximizing the societal incentive power amounts to solving
\begin{equation}\label{jointproblem}
\max_{\{a_i,b_i\}_{i=1}^8} \left(\ref{joint}\right) \text{ s.t. } (\ref{feasibility}) \text{ and } (\ref{consistency}).
\end{equation}
As demonstrated by the next example, the solution to  (\ref{jointproblem}) could exhibit correlations between different voters' signals. 

\begin{example}\label{exm_correlation}
In the case where $I_k =.1$ $\forall k \in \mathcal{K}$, $v_1=.24$, and $h\left(x\right)=x^2$, solving (\ref{jointproblem}) numerically yields  $a_1=0$, $a_2=.024$, $a_3=.296$, $a_4=.633$, $a_5=0$, $a_6=0$, $a_7=0$, $a_8=.0456$, $b_1=.0456$, $b_2=0$, $b_3=0$, $b_4=0$, $b_5=.633$, $b_6=.296$, $b_7=.024$, and $b_8=0$. The societal incentive power equals $.908$ and is greater than the societal incentive power $.457$ generated by the joint distribution whereby signals are conditionally independent across voters.  $\hfill \diamondsuit$
\end{example}

\section*{Declaration of competing interest}
None.

\section*{Acknowledgments}
We thank Ken Shotts for his encouragement and inputs throughout the development of the project,  Justin Fox for insightful feedback on an earlier draft, and the seminar audience at Olin Business School for interesting discussions.

\begin{spacing}{1}
\bibliographystyle{econometrica} 
\bibliography{ri-accountability.bib}
\end{spacing}

\end{document}